\documentclass{jfm}
\usepackage{graphicx}
\usepackage{epstopdf, epsfig}
\usepackage{cancel}
\usepackage{multirow} 
\usepackage{amsmath}
\usepackage{array}
\usepackage{amssymb}
\usepackage{natbib} 		
\usepackage{hyperref}
\usepackage{color}
\usepackage{pbox}
\usepackage{subcaption}
\usepackage{tikz}
\usepackage{floatrow}
\usepackage{mathrsfs}
\usepackage[utf8]{inputenc}

\newcolumntype{M}[1]{>{\centering\arraybackslash}m{#1}}

\newcommand{\tikzcircle}[2][black,fill=black]{\tikz[baseline=-0.5ex]\draw[#1,radius=#2] (0,0) circle ;}

\shorttitle{Thermoacoustic shock waves}
\shortauthor{P. Gupta, G. Lodato, and C. Scalo}

\title{Spectral energy cascade in thermoacoustic shock waves}

\author{Prateek Gupta\aff{1}
  \corresp{\email{gupta288@purdue.edu}},
  Guido Lodato\aff{2}
 \and Carlo Scalo\aff{1}}

\affiliation{\aff{1}School of Mechanical Engineering, Purdue University,
West Lafayette, IN 47906, USA
\aff{2} Normandie Universit{\'e}, CNRS, INSA et Universit{\'e} de Rouen, CORIA UMR6614, France}

\begin{document}

\maketitle

\begin{abstract} 

We have investigated thermoacoustically amplified quasi-planar nonlinear waves driven to the limit of shock-wave formation in a variable-area looped resonator geometrically optimized to maximize the growth rate of the quasi-travelling-wave second harmonic. Optimal conditions result in velocity leading pressure by approximately $40^\circ{}$ in the thermoacoustic core and \emph{not} in pure travelling-wave phasing. High-order unstructured fully compressible Navier-Stokes simulations reveal three regimes: ($i$) \emph{Modal growth}, governed by linear thermoacoustics; ($ii$) \emph{Hierarchical spectral broadening}, resulting in a nonlinear inertial energy cascade; $(iii)$ \emph{Shock-wave dominated limit cycle}, where energy production is balanced by dissipation occurring at the captured shock-thickness scale. The acoustic energy budgets in regime ($i$) have been analytically derived, yielding an expression of the Rayleigh index in closed form and elucidating the effect of geometry and hot-to-cold temperature ratio on growth rates. A time-domain nonlinear dynamical model is formulated for regime ($ii$), highlighting the role of second-order interactions between pressure and heat-release fluctuations, causing asymmetry in the thermoacoustic energy production cycle and growth rate saturation. Moreover, energy cascade is inviscid due to steepening in regime ($ii$), with the $k^{\mathrm{th}}$ harmonic growing at $k/2$-times the modal growth rate of the thermoacoustically sustained second harmonic. The frequency energy spectrum in regime ($iii$) is shown to scale with a $-5/2$ power law in the inertial range, rolling off at the captured shock-thickness scale in the dissipation range. We have thus shown the existence of equilibrium \emph{thermoacoustic energy cascade} analogous to hydrodynamic turbulence.
 
\end{abstract}

\begin{keywords}
To be given online
\end{keywords}


\section{Introduction}

Thermoacoustic amplification of waves in a compressible flow is the result of a fluid dynamic instability emerging from the favourable coupling between pressure and heat-release fluctuations. The wavelength of thermoacoustically unstable waves is set by the size of the enclosing resonant chamber, while the heat release, providing the energy source for the amplification, is confined to a compact region. The heat release rate is a function of local velocity and pressure fluctuations, affecting for example the instantaneous flame surface area in a combustion chamber or the rate of convective heat extraction from a hot wire-mesh screen in a Rijke tube. The resulting fluctuations in the heat release rate drive a cycle of compressions and dilatations, which act as a source for pressure fluctuations that, if within a quarter phase from the heat release itself, become thermoacoustically amplified (\citealt{Rayleigh_Nature_1878}'s criterion). In combustion applications, efforts are made to suppress thermoacoustic amplification of waves~\citep{PoinsotV_numComb_2011}, which is, on the other hand, desired in thermoacoustic waste heat recovery devices~\citep{Swift_1988_JAcoustSA}.

The goal of the present manuscript is to investigate the high amplitude (or macrosonic) limit of thermoacoustically driven nonlinear waves characterized by the formation of self-sustaining resonating shock waves and inter-harmonic, or inter-scale, energy transfer. In thermoacoustically unstable combustors, modelling based on a modally truncated analysis of nonlinearly coupled oscillators has been shown to successfully describe the energy transfer among harmonics~\citep{Culik_AGARD_2006, Ananthkrishnan_CST_2005}.~\cite{Culick_CST_1971} also formalized the procedure for estimating the nonlinear growth and decay of the thermoacoustically unstable mode and a few companion overtones (modally truncated analysis). However, in the limit of self-sustaining resonating shock waves, harmonics with very short wavelengths are generated as a result of energy cascade, rendering modally truncated analysis insufficient. While the problem of shock wave resonance sustained by near-resonant frequency excitation has been analysed in detail~\citep{Saenger_JASA_1960, Chester_JFM_1964}, self-sustaining shocks generated via modal instability remain to be analysed in detail to the best of authors' knowledge.

Multiple authors~\citep{Swift_JAcoustSocAm_1992,KarpovP_JAcoustSocAm_2000,HamiltonIZ_JAcoustSocAm_2002,PeneletGLB_PhysRevE_2005} have discussed the higher harmonic generation in non-combustion-driven thermoacoustically unstable waves, restricting their analysis to only a first few overtones of the unstable mode.~\cite{BiwaTY_JAcousSocAm_2011} reported the first experimental observation of energy cascade in thermoacoustically sustained shock waves in a looped thermoacoustic resonator. The stability limits of a similar experimental setup were studied previously by~\cite{Yazaki_PhysRevLet_1998}, who generated second-mode quasi-travelling wave instability, although no shock formation was reported.~\cite{olivier_weakly_2015} performed numerical modelling of a looped thermoacoustic resonator similar (but not identical to)~\cite{BiwaTY_JAcousSocAm_2011}'s apparatus, combining approximate nonlinear propagation equations~\citep{MenguyG_AAA_2000} in the resonator with linear thermoviscous equations~\citep{Rott_ZAMP_1969,Rott_ZAMP_1973} in the thermoacoustic regenerator, i.e. the heat exchanger used to sustain the wall temperature gradient required to generate the instability. The approximate equations by~\cite{MenguyG_AAA_2000}, however, assume low frequencies, hence are invalid over the entire spectrum generated by a nonlinear energy cascade driven up to shock wave formation, and neglect thermoacoustic nonlinearities in the regenerator, which, as shown later, need to be accounted for.

In the present work, a comprehensive nonlinear theoretical and high-fidelity modelling approach is adopted to accurately describe macrosconic thermoacoustic waves. To this end, a canonical travelling-wave looped resonator, inspired by~\cite{Yazaki_PhysRevLet_1998}'s experimental setup but geometrically optimized via linear theory~\citep{Rott_ZAMP_1969,Rott_ZAMP_1973,LinSH_JFM_2016}, has been designed to maximize the growth rate of the quasi-travelling-wave second harmonic and thus achieve rapid shock wave formation.~\cite{Yazaki_PhysRevLet_1998}'s looped configuration allows quasi-travelling-wave acoustic phasing which facilitates faster nonlinear energy cascade compared to standing wave resonators~\citep{BiwaEtAl_JASA_2014}. It is shown that the energy content in spectral domain resembles the equilibrium energy cascade observed in turbulence, similar to the spectral energy distribution of an ensemble of acoustic waves interacting nonlinearly among each other~\citep{Nazarenko_2011_WT, Zakharov_2012_WT}. As demonstrated by the numerical simulation data and companion low-order nonlinear modelling, thermoacoustically sustained shock waves exhibit inter-scale energy transfer dynamics analogous to Kolmogorov's equilibrium hydrodynamic turbulent energy cascade~\citep{kolmogorov1941local}. Throughout, the results obtained via the proposed  nonlinear model are verified and compared with fully compressible high-fidelity Navier-Stokes simulations.

The development of an accurate nonlinear thermoacoustic wave propagation theory and modelling framework warrants the support of high-fidelity numerical simulations. A high-order spectral difference numerical framework~\citep{kopriva1996conservative,kopriva1996conservative2,sun2007high,Jameson_JSC_2010} combined with an artificial Laplacian viscosity~\citep{persson:06} shock-capturing scheme has been adopted for the present study. Moreover, the computational setup has been reduced to a minimal-unit (or single-pore) configuration, as done by~\cite{RahmanEtAl_IntJHMT_2017}, to reduce the computational cost and ensure the maximum possible numerical resolution in the direction of shock propagation for a given number of discretization points, or degrees of (numerical) freedom. In spite of this choice, full resolution of the propagating shocks was still not attainable with the available resources.
 
The paper is organized as follows. Details of the computational setup are presented in \S~\ref{sec: ComputationalSetup}, followed by a grid convergence study. In \S~\ref{sec: Regimes}, the various regimes of thermoacoustic amplification and energy cascade are identified. In \S~\ref{sec: Linear}, the dependence of instability growth rates on acoustic phasing is presented utilizing the acoustic energy budgets. A time-domain nonlinear acoustic model is developed in \S~\ref{sec: NonlinearRegime_modelling}, highlighting the importance of accounting for thermodynamic nonlinearities to match the results from the Navier-Stokes simulations presented in \S~\ref{sec: nonlinear_cascade}. Finally, in \S~\ref{sec: LimitCycle}, the distribution of the spectral energy density at steady state is discussed and modelled heuristically based on dimensional analysis statements, establishing the existence of an equilibrium \emph{thermoacoustic energy cascade} analogous to hydrodynamic turbulent cascade.


\section{Problem Formulation} \label{sec: ComputationalSetup}

\subsection{Design of the minimal-unit (or single-pore) thermoacoustic resonator model}
\label{sec: optimization}

\begin{figure}
\centering
\includegraphics[width=0.85\textwidth]{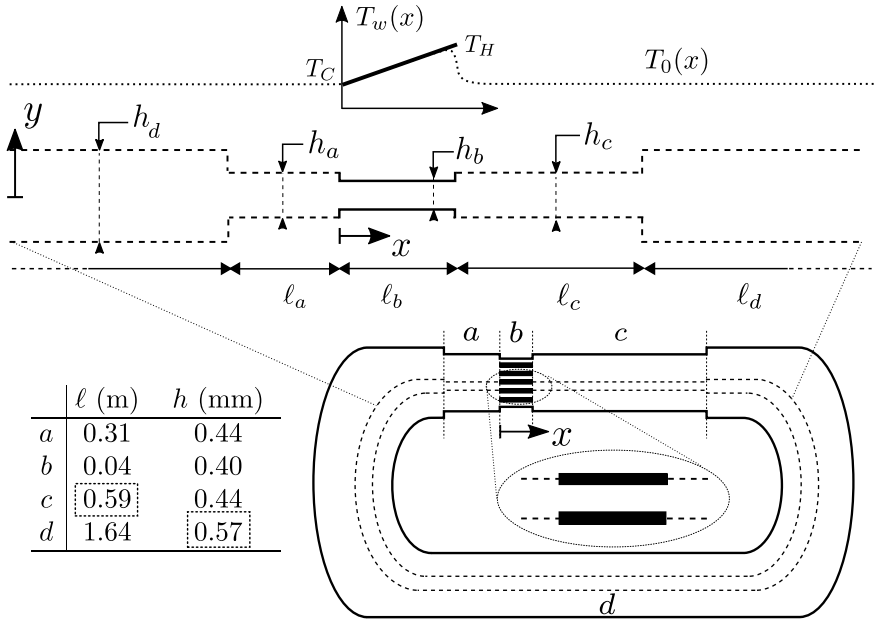}
  \caption{Two-dimensional axially periodic computational setup for minimal-unit simulations (top, not to scale), geometrical parameters (bottom left), and qualitative illustration of the equivalent experimental build-up of a variable-area looped thermoacoustic resonator (bottom right). Top figure: (-- --), adiabatic/sip conditions; (--), isothermal/no-slip conditions. The minimal unit is traced in the bottom right with dashed lines (-- --). The listed geometrical parameters provide sub-optimal growth rates across all the values of $T_H$, the boxed quantities have been determined via optimization (cf.~figure~\ref{fig:optimization_lst} and table~\ref{tbl:base_state}).}
\label{fig:computational_setup}
\end{figure}

The proposed computational setup (figure~\ref{fig:computational_setup}, top) is a straight, two-dimensional, axially periodic minimal-unit (or single-pore)  thermoacoustic device composed of four constant-area sections ($a$, $b$, $c$, and $d$). Such configuration represents an idealization of a looped thermoacoustic resonator (figure~\ref{fig:computational_setup}, bottom) similar---but not identical---to the one adopted by~\cite{Yazaki_PhysRevLet_1998}. Adiabatic slip conditions are applied everywhere, with the exception of the thermoacoustic regenerator, or core (section $b$), where isothermal no-slip walls are used to impose a linear wall-temperature distribution $T_w(x)$ from the cold, $T_C$, to the hot side temperature, $T_H$, resulting in the base temperature distribution, $T_0(x)$. A body force is added to suppress Gedeon streaming (\S~\ref{sec:gov_equations}), which would otherwise cause convective heat transport away from the hot end of section $b$ and require the introduction of a thermal buffer tube and a secondary ambient heat exchanger~\citep{PeneletGLB_PhysRevE_2005}, thus introducing further complications in the proposed canonical setup (discussed in \S~\ref{sec:gov_equations}). The resulting relaxation of base state from $T_H$ to $T_C$ outside the regenerator is due to molecular diffusion. Consequently, a diffusive thermal layer develops which is very thin compared to the acoustic wave length and is neglected in the subsequent analysis.
The uniform base pressure and cold-side values of density and temperature are set to be equal to the reference thermodynamic quantities $P_0=P_{\mathrm{ref}}$, $\rho_C=\rho_{\mathrm{ref}}$, $T_C=T_{\mathrm{ref}}$ (table~\ref{tbl:base_state}), chosen for air (\S~\ref{sec:gov_equations}). Stacking of any number of thus-conceived single-pore models in the $y$ direction, i.e. preserving the area ratio and increasing the number of pores in the regenerator, would yield the same numerical results. 

The minimal-unit choice is dictated by the need to maximize the numerical resolution in the propagation direction of the captured shocks. Although full resolution of the propagating shocks is still not computationally feasible due to the very large length of the setup (of the order of $1$\,m) compared to the typical shock thickness scale (of the order of $100\,\mu$m). Performing a fully resolved three-dimensional simulation of an equivalent experimental setup would be (even more so) unfeasible: for instance, the setup studied by~\cite{Yazaki_PhysRevLet_1998} consists of two heat exchangers and a ceramic catalyst with approximately 1000 pores. By design, the minimal-unit configuration neglects the thermoviscous losses outside the regenerator that attenuate the thermoacoustic amplification. The two-dimensional effects of the curvature of the resonator walls are also neglected. The former become less important as the ratio of the pore diameter to resonator diameter and number of pores increases (42 and 1000, respectively, in Yazaki's setup), and the latter are only relevant for very small ratios of the curvature radius to acoustic wavelength ($\sim 0.3$ in Yazaki's setup, assuming curvature radius of $L/6$ where $L$ is the total length). In conclusion, performing a full three-dimensional simulation taking into account approximately 1000 thermoacoustic pores, viscous losses in the resonator, and the curvature effects, would not lead to any significant additional insights into the physics of thermoacoustically generated shock waves, especially within the theoretical scope of the present study.

\begin{table}
  \begin{tabular}{M{1.7cm}M{2.0cm}M{1.3cm}M{2.4cm}}
 \hline
$P_0$ & $\rho_C$ & $T_C$ & $T_H$ \\
 \multirow{2}{*}{101325\,Pa} & \multirow{2}{*}{1.176 kg/m$^3$} & \multirow{2}{*}{300\,K} & 400\,K, 450\,K, \\
&   &  & 500\,K, 550\,K \\
  \hline
\end{tabular}
 \caption{Thermodynamic parameters for base state. Base pressure and cold-side values of density and temperature are set to reference values: $P_0=P_{\mathrm{ref}}$, $\rho_{c}=\rho_{\mathrm{ref}}$, $T_C=T_{\mathrm{ref}}$.}
 \label{tbl:base_state}
 \end{table}

\begin{figure}
\centering
\includegraphics[width=\textwidth]{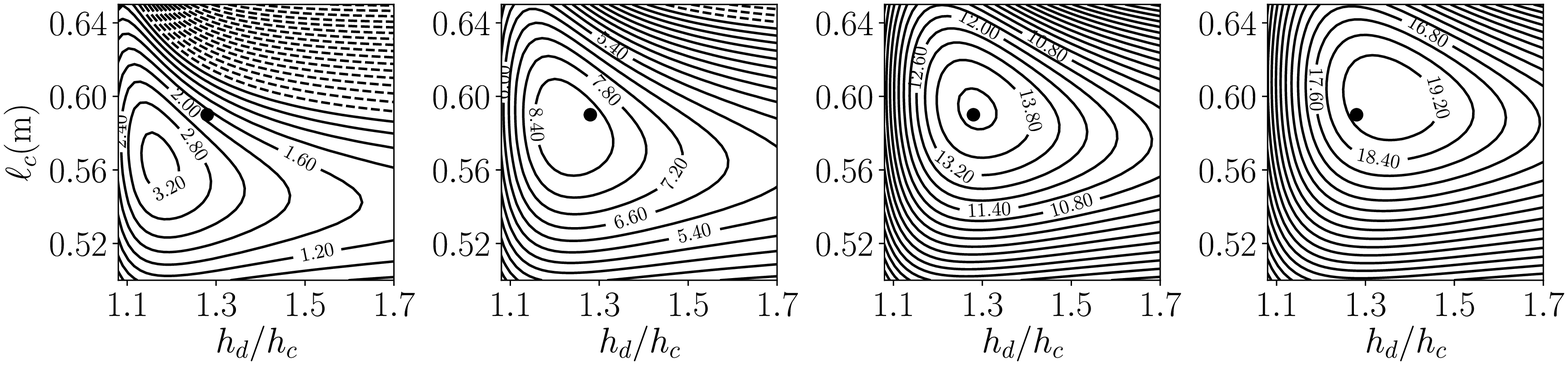}
\put(-370,5){$(a)$}
\put(-275, 5){$(b)$}
\put(-185,5){$(c)$}
\put(-90,5){$(d)$}\\
  \begin{tabular}{M{1cm}M{1.4cm}M{1.4cm}M{1.4cm}M{1.4cm}}
 \hline
$T_H$ & 400\,K & 450 \,K & 500\,K & 550\,K \\
\rule{0pt}{1ex}
$\alpha$ & 1.79\,$\mathrm{s}^{-1}$ & 8.64\,$\mathrm{s}^{-1}$ & 14.51\,$\mathrm{s}^{-1}$ & 19.43\,$\mathrm{s}^{-1}$\\
  \hline
\end{tabular}
  \caption{Iso-contours of thermoacoustic growth rate, $\alpha$, of the second harmonic versus resonator area ratio, $h_d/h_c$, and length $\ell_c$ (see figure~\ref{fig:computational_setup}) for hot-side temperatures $T_H=400~\mathrm{K}$ ($a$), $T_H=450~\mathrm{K}$ ($b$), $T_H=500~\mathrm{K}$ ($c$), and $T_H=550~\mathrm{K}$ ($d$). (--), $\alpha > 0$; (- -), $\alpha < 0$; (\tikzcircle{2pt}), sub-optimal values $\ell_c = 0.59$ m and $h_{d}/h_{c} = 1.28$ chosen for present investigation (figure~\ref{fig:computational_setup}). Growth rate values for each $T_H$ at sub-optimal geometry are listed in the table.}
\label{fig:optimization_lst}
\label{tbl:growth_rates}
\end{figure}

The total length of the device is fixed to $\ell_a + \ell_b + \ell_c + \ell_d = 2.58~\mathrm{m}$, taken from the experimental setup of~\cite{Yazaki_PhysRevLet_1998}. The height of the regenerator has been chosen such that $h_b\sim2\delta_k$, where
\begin{equation}
\delta_k = \sqrt{\frac{2\nu}{\omega Pr}},
\end{equation}
$\nu$ is the kinematic viscosity, $Pr$ is the Prandtl number, and $\omega$ is the angular frequency of the unstable mode.
This results in a porosity $h_b/h_c=0.91$. The area ratio $h_d/h_c$ and the length $\ell_c$ have been chosen to yield sub-optimal values of growth rates across all hot-side temperature settings (figure~\ref{fig:optimization_lst}), assuring high enough thermoacoustic instability to achieve rapid shock wave formation. This optimization has been carried out with the system-wide numerical approach developed by~\cite{LinSH_JFM_2016}, using the  governing equations outlined in \S~\ref{sec: LinearRegime}. Further justification of the setup choice and a comparison of the linear numerical analysis with~\citet{Yazaki_PhysRevLet_1998}'s stability data are given in appendix~\ref{sec: app0}.

\subsection{Navier-Stokes calculations}

\subsubsection{Governing Equations} \label{sec:gov_equations}
Fully compressible Navier-Stokes simulations are carried out by solving the conservation laws for mass, momentum, and total energy in two dimensions, given by
\begin{subequations}
	\label{eq:navierstokes}
	\begin{align}
		\frac{\partial}{\partial t} \left(\rho\right) &+ \frac{\partial}{\partial x_j} \left(\rho u_j \right)  = 0,
		\label{subeq:ns1}
		\\
		\frac{\partial}{\partial t} \left(\rho u_i\right) &+ \frac{\partial}{\partial x_j} \left(\rho u_i u_j\right)  =  -\frac{\partial}{\partial x_i} p  +
		\frac{\partial}{\partial x_j} \tau_{ij} + \delta_{1i} f_{B},
		\label{subeq:ns2}
		\\
		\frac{\partial}{\partial t} \left(\rho \, E\right) &+ \frac{\partial}{\partial x_j} \left[ u_j \left(\rho \, E + p \right) \right] =
		\frac{\partial}{\partial x_j } \left(u_i \tau_{ij} - q_j\right),
		\label{subeq:ns3}
	\end{align}
\end{subequations}
respectively, where $x_i$($x_1$, $x_2$ or equivalently, $x$, $y$) are the axial and cross-sectional coordinates, $u_i$ are the velocity components in each of those directions, and $p$, $\rho$, $T$, and $E$ are the instantaneous pressure, density, temperature, and total energy per unit mass, respectively. 

Due to the propagation of finite amplitude nonlinear acoustic waves and the periodic nature of the setup, Gedeon streaming~\citep{Gedeon_Cryo_1995} is expected. The mean flow caused by the Gedeon streaming results in the transport of heat away from the hot side of the regenerator into the whole device affecting the mean temperature distribution outside the regenerator. Such thermal leakage is usually mitigated by a secondary cold heat exchanger and a thermal buffer tube to achieve steady-state conditions. Due to the inhomogeneous mean temperature outside the regenerator, such a setup would exhibit a very large design parameter space. Moreover, the mean flow caused by the streaming would further disperse the acoustic waves, affecting the wave steepening and thus delaying the steady state further. Such effects fall beyond the scope of the present investigation. Thus, acoustic streaming in the current work is purposefully suppressed. To this end, a uniform mean pressure gradient $f_{B}$ in the axial direction is dynamically adjusted to relax the net axial mass flow rate to zero. The relevant expression for $f_{B}$ is: 
\begin{equation}
f_B = \frac{\alpha}{\Delta t}\left(\dot{m} - \dot{m}_0\right), 
\end{equation}
where $\alpha = 0.3$ is a relaxation coefficient, $\Delta t$ is the time step, $\dot{m}$ is the instantaneous volume averaged mass flow rate, and $\dot{m}_0 = 0$ is the target value. The viscous stress tensor, $\tau_{ij}$, and heat flux, $q_{i}$, are formulated based on the Stokes and the Fourier laws as
\begin{subequations}
\label{eq:heatfluxes}
  \begin{eqnarray}
		\tau_{ij} = 2 \mu \left[S_{ij} - \frac{1}{3} \frac{\partial u_k}{\partial x_k} \delta_{ij} \right],\quad q_j &=& -\frac{\mu\,C_p}{Pr} \frac{\partial T}{\partial x_j} ,
		\label{subeq:hf2}
  \end{eqnarray}
\end{subequations}
respectively, where $S_{ij}$ is the strain-rate tensor, given by $S_{ij}=\frac{1}{2} \left(\partial u_j/\partial x_i + \partial u_i /\partial x_j \right)$, $Pr$ is the Prandtl number, and $C_p$ is the specific heat capacity at constant pressure. The working fluid is air, assumed to be calorically and thermally perfect. The dynamic viscosity, $\mu(T)$, is varied with the temperature according to the Sutherland's law, $\mu(T) = \mu_\textrm{ref}(T/T_{S,\textrm{ref}})^{1.5}(T_{S,\textrm{ref}}+S)/(T+S)$ where $S = 120$ K is the Sutherland constant, $T_{S,\textrm{ref}} = 291.15$ K and $\mu_\textrm{ref} = 1.827\times10^{-5}$~\,kg$\cdot$m$^{-1}\cdot$s$^{-1}$. The values of the other unspecified fluid parameters, valid for air, are $Pr = 0.72$, the ratio of isobaric to isochoric specific heat capacities $\gamma=1.4$, the reference density $\rho_{\textrm{ref}} = 1.176$~\,kg$\cdot$m$^{-3}$, the pressure $p_{\textrm{ref}} = 101\,325$~\,Pa, the temperature $T_{\textrm{ref}}=300$~\,K, and the gas constant $R=p_{\textrm{ref}}/(\rho_{\textrm{ref}}\,T_{\textrm{ref}})$.
\begin{figure}
  \centerline{\includegraphics[width=1.0\textwidth]{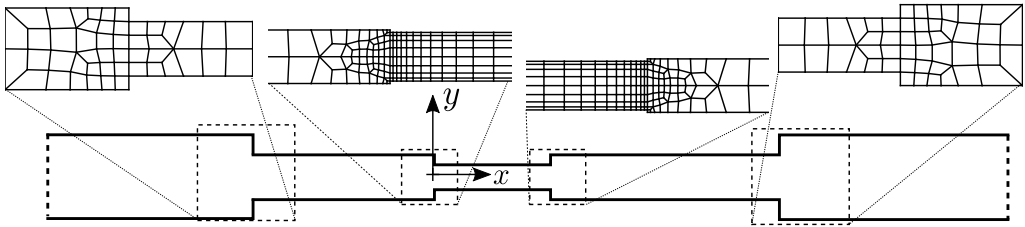}}
  \caption{Illustration of selection portions of the unstructured mesh near duct interfaces for Grid A with total number of elements $N_{\mathrm{el}}=2004$.}
\label{fig:mesh}
\end{figure}

\begin{table}
\centering
\hspace*{-0.2cm}
\begin{tabular}{c}
\multicolumn{1}{c}{} \\
\rule{0pt}{4.9ex} 
\vspace*{-0.18cm}
\\
$p=3$\\
$p=5$ \\
$p=7$ \\
\end{tabular}
\begin{tabular}{ccc}
\multicolumn{3}{c}{Grid A ($N_{\mathrm{el}}=2004$) } \\
\rule{0pt}{3.8ex}  
$N_{\mathrm{dof}}$ & ${\Delta_h}_{\mathrm{(min)}}$ & ${\Delta_h}_{\mathrm{(max)}}$ \\
\hline
18036 & 7.33 $\mu$m & 0.63 mm\\
50100 & 4.40 $\mu$m & 0.37 mm\\
98196 & 3.14 $\mu$m & 0.26 mm\\
\hline
\end{tabular}\quad\vline\quad
\begin{tabular}{ccc}
\multicolumn{3}{c}{Grid B ($N_{\mathrm{el}}=7087$) } \\
\rule{0pt}{3.8ex}  
$N_{\mathrm{dof}}$ & ${\Delta_h}_{\mathrm{(min)}}$ & ${\Delta_h}_{\mathrm{(max)}}$ \\
\hline
63783 & 3.67 $\mu$m & 0.52 mm\\
177175 & 2.20 $\mu$m & 0.31 mm\\
347263 & 1.57 $\mu$m & 0.22 mm\\
\hline
\end{tabular}\quad
 \caption{Discretization order used per element, $p$, total number of elements, $N_{\mathrm{el}}$, degrees of freedom, $N_{\mathrm{dof}} = p^2 N_{\mathrm{el}}$, and minimum and maximum linear element size ${\Delta_h}_{\mathrm{(min)}}$, ${\Delta_h}_{\mathrm{(max)}}$ (cf.~(\ref{eq:linear_refinement_ratio})). Reported values are the ones used at the limit cycle. In all cases, simulations are carried out with $p=3$ and Grid A throughout the transient.}
 \label{tab:GridParams}
\end{table}
The fully compressible high-fidelity Navier-Stokes calculations have been carried out with the discontinuous finite element \textsc{sd3DvisP} solver, an MPI parallelized Fortran 90 code employing the spectral difference local spatial reconstruction for hexahedral elements on unstructured grids~\citep{kopriva1996conservative,kopriva1996conservative2,sun2007high,Jameson_JSC_2010}. The solver reconstructs the local solution inside each element as the tensor product of polynomials up to the user-specified order $p = N_p-1$, where $N_p$ is the number of solution points per dimension inside the element. Inter-element discontinuities in the solution are handled utilizing the~\cite{roe1981approximate}'s flux with the entropy correction by~\cite{harten1983self}. The numerical dissipation at the element interfaces scales as $\Delta_h^{N_p+1}$ where $\Delta_h$ is the characteristic length scale of neighbouring elements~\citep{lodato:14b,ChapelierLJ_CF_2016}. Sub-cell shock capturing is enabled through a Laplacian artificial diffusion term applied in regions of steep gradients, which are detected by means of a modal sensor based on a Legendre polynomial expansion~\citep{persson:06,lodato:16}. The time integration is carried out explicitly with a 3rd order Runge-Kutta scheme and discretization order $p=3$ during the transient, whereas a 5th order Runge-Kutta scheme is adopted at the shock-dominated limit cycle where a systematic grid sensitivity study has been carried out (cf.~table~\ref{tab:GridParams} and figure~\ref{fig:GridConv}), as discussed in the following section. The same solver has been used and validated in a wide variety of flow configurations, including turbulent channel flow~\citep{lodato2013discrete, lodato2014structural} and unsteady shock-wavy wall interaction problems~\citep{lodato:16,lodato:17}. 

\subsubsection{Grid Sensitivity Analysis}
\begin{figure}
\centering
\includegraphics[width=1.0\textwidth]{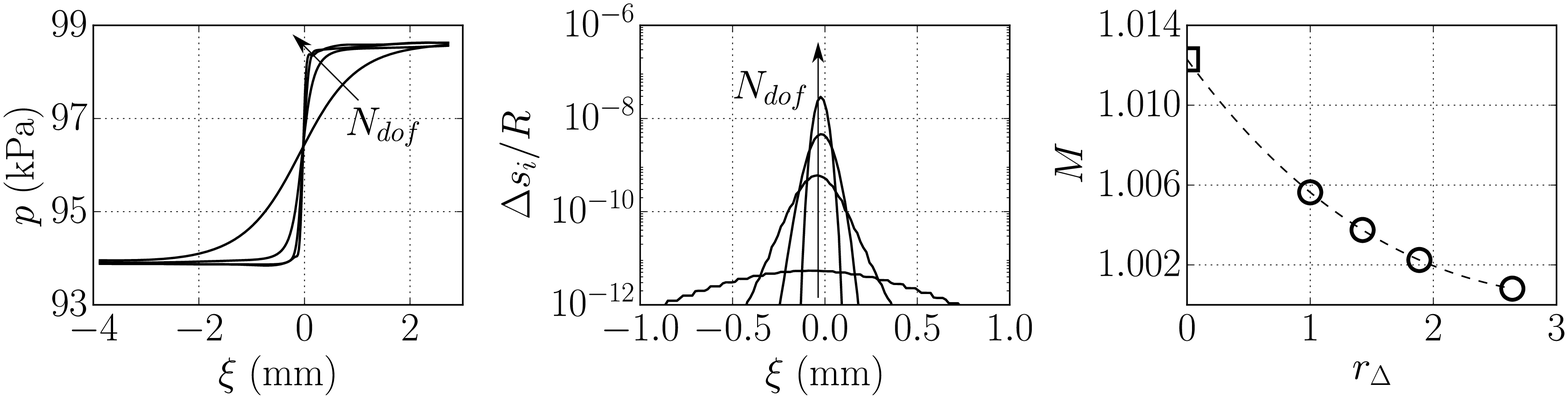}
\put(-385,5){$(a)$}
\put(-255,5){$(b)$}
\put(-110,5){$(c)$}
  \caption{Grid sensitivity analysis at the shock dominated limit cycle for $T_H=450$K. Pressure versus travelling-wave coordinate $\xi$ for increasing numerical degrees of freedom (table~\ref{tab:GridParams}) from time series extracted at $x=-2$ mm (figure~\ref{fig:computational_setup}) $(a)$,  point-to-point dimensionless entropy increments $(b)$, and shock Mach number $M$ versus grid refinement ratio $r_{\Delta}$ $(c)$. ($c$): (- -), Polynomial extrapolation (cf.~\eqref{eq: Richardson}); ($\circ{}$), shock Mach number; ($\square$), Mach number in the continuum limit $r_{\Delta}\rightarrow 0$.}
\label{fig:GridConv}
\end{figure}
A grid sensitivity analysis is carried out at the shock-dominated limit cycle spanning the simulation parameter space outlined in table~\ref{tab:GridParams} only for the $T_H = 450$ K temperature setting. The pressure waveform, expressed as a function of the travelling wave coordinate
\begin{equation} \label{eq:wave_xsi}
\xi = c_s\,t - x,
\end{equation}
where $c_s \simeq a_0$ is the (weak) shock speed, and $a_0$ is the base speed of sound, monotonically converges to a step function (figure~\ref{fig:GridConv}$a$) as the numerical degrees of freedom, $N_{\mathrm{dof}}$, of the high-order Navier-Stokes calculations are increased. The Mach number of the captured shock waves (figure~\ref{fig:GridConv}$c$) is estimated, for each grid resolution level in table~\ref{tab:GridParams}, by calculating the total entropy jump, $\Delta s$, as the sum of the point-to-point entropy increments: 
\begin{equation}
\frac{\Delta s}{R} = \sum_i \frac{\Delta s_i}{R} \approx \frac{\gamma + 1}{12\gamma^2}\sum_i \left(\frac{\Delta p_i}{p_i}\right)^3, \label{eq: entropy_jump}
\end{equation}
where $\Delta p_i = p_{i+1}-p_i$, $\Delta s_i = s_{i+1}-s_i$ are the pressure and the entropy jumps from point $i$ to $i+1$ in the $\xi$ space (figure~\ref{fig:GridConv}$b$). Thus the Mach number of the captured wave is obtained upon solving 
\begin{equation}
 \frac{\Delta s}{R} \approx \frac{2\gamma}{(\gamma+1)^2}\frac{(M^2-1)^3}{3}.
\end{equation}
Evaluating the overall entropy jump, and therefore, the shock Mach number, based on the total pressure jump, would include spurious contributions from isentropic compression waves which do not participate in the coalescence of waves forming the shock. Finally, the shock Mach number $M_{r_{\Delta}\rightarrow0}$ at continuum limit is estimated via a polynomial extrapolation,
\begin{equation}
 M(r_{\Delta}) = m_0 + m_1\,r_{\Delta} + m_2\,r_{\Delta}^2 + m_3\,r_{\Delta}^3,\label{eq: Richardson}
\end{equation}
where $m_0$, $m_1$, $m_2$, and $m_3$ are fitting parameters and $r_{\Delta}$ is the grid refinement ratio,
\begin{equation} \label{eq:linear_refinement_ratio}
r_{\Delta} = \overline{\Delta}_h/\overline{\Delta}_h|^B_{p=7},~~\overline{\Delta}_h = \sqrt{\Omega/N_{\mathrm{dof}}},
\end{equation}
in which $\Omega$ denotes total area of the computational domain.


\section{Regimes of Thermoacoustic Amplification}
\label{sec: Regimes}

\begin{figure}
\centering
  \centerline{\includegraphics[width=\textwidth]{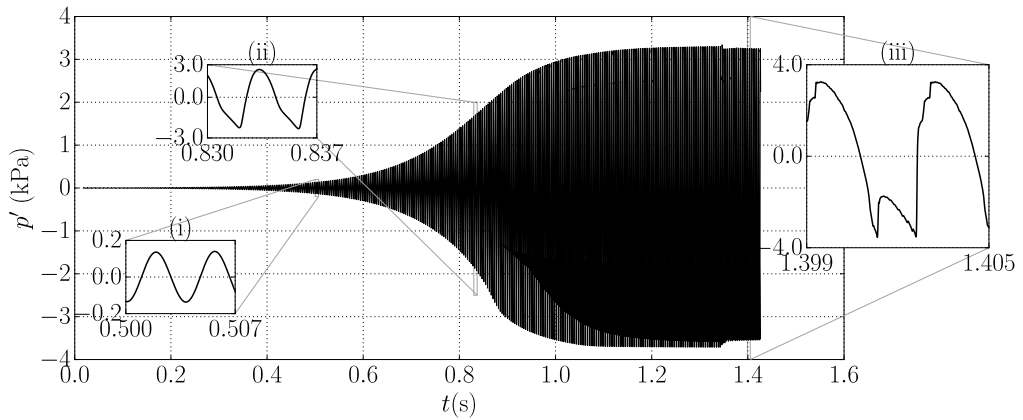}}
  \caption{Time series of pressure fluctuations at $x=1.54$ m (figure~\ref{fig:computational_setup}) for $T_H=$450 K with insets showing regimes ($i$) modal growth, ($ii$) hierarchical spectral broadening, and ($iii$) the shock-dominated limit cycle.}
\label{fig: Intro_signal}
\end{figure}

Three different regimes of thermoacoustic wave amplification can be identified by visual inspection of the pressure time series in figure~\ref{fig: Intro_signal}. We attempt a rigorous classification here based on the dimensionless collapse of the nonlinear growth regime of the spectral energy density (figure~\ref{fig:intro_spectral}), derived in more detail in \S~\ref{sec: Spectral energy distribution}.

\subsection{Spectral energy density}
Any acoustic dynamical model (linear or nonlinear) can be written as 
\begin{equation}
\frac{\partial \mathbf{X}}{\partial t} = \mathbf{f}(\mathbf{X}),~~\mathrm{with}~~\mathbf{X} = \left(\frac{u'}{a_0}, \frac{p'}{\rho_0a^2_0}\right)^{\mathrm{T}},
\label{eq: AcousticDynamicalSystem}
\end{equation}
where $\mathbf{X}$ is the state vector containing the dimensionless perturbation variables. Throughout, we define the following squared $L_2$ norm as the perturbation energy density~\citep{Naugol1975spectrum}:
\begin{equation}
E = \frac{1}{2}\rho_0a^2_0 \mathbf{X}^{\mathrm{T}}\mathbf{X} = \frac{1}{2}\rho_0u'^2 + \frac{p'^2}{2\rho_0a^2_0}.
\label{eq:energy_time}
\end{equation}
In the nonlinear growth and limit cycle regimes, velocity and pressure fluctuations, $u'$ and $p'$, respectively, are composed of higher harmonics of the linearly unstable mode. Substituting velocity and pressure fluctuations, expressed as complex Fourier expansions, 
\begin{align}
u'(x,t) = \sum_{\substack{k=-\infty\\k\ne0}}^{+\infty}\hat{u}_k(x,\epsilon t)\,e^{i\frac{k}{2}\omega_0t} ,& \quad p'(x,t) = \sum_{\substack{k=-\infty \\k\ne0}}^{+\infty} \hat{p}_k(x,\epsilon t)e^{i\frac{k}{2}\omega_0t},\\
\hat{u}_{-k} = \hat{u}^{*}_{k},& \quad \hat{p}_{-k} = \hat{p}^{*}_{k}, \nonumber
\end{align}
into~\eqref{eq:energy_time} and cycle averaging yields: 
\begin{equation}
\overline{E} = 2\sum_{k=1}^{\infty} E_k, \quad E_k = \frac{1}{2}\rho_0|\hat{u}_k|^2 + \frac{|\hat{p}_k|^2}{2\rho_0a^2_0}, 
\label{eq:energy_spectral}
\end{equation}
where $\epsilon\sim \alpha/\omega \ll 1$ is the smallness parameter such that $t$ and $\epsilon t$ correspond to fast and slow time scales, respectively, $\overline{(\cdot)}$ denotes the cycle averaging operator defined as
\begin{equation}
\overline{(\cdot)} = \frac{1}{T_0}\int_{\epsilon t}^{\epsilon t+T_0} (\cdot)dt, \quad T_0=\frac{2\pi}{\omega_0},
\end{equation}
and $(\cdot)^*$ denotes the complex conjugate. Here $\omega_0$ is the angular frequency of the unstable second harmonic, and $E_k$ is the spectral energy density of the $k^{\mathrm{th}}$ mode. The pressure and velocity amplitudes of the $k^{\mathrm{th}}$ harmonic ($|\hat{p}_k|$ and $|\hat{u}_k|$, respectively) are functions of the $x$ coordinate and the slow time $\epsilon t$, and are extracted via a short time-windowed Fourier transform (over 8 cycles of time period $T_0$) of the time series shown in figure~\ref{fig: Intro_signal}. In the nonlinear growth regime, the energy cascades from the unstable second mode ($k=2$) into its overtones only ($k=4, 6, 8, \dots$) with no energy content in the odd-numbered harmonics. 

\subsection{Regime classification}
\begin{figure}
\centering
  \centerline{\includegraphics[width=0.9\textwidth]{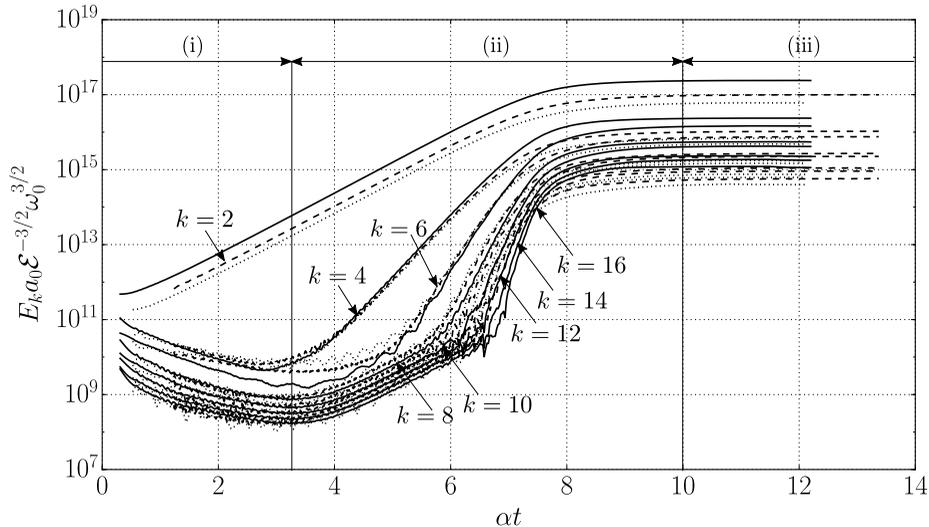}}
  \caption{Evolution of dimensionless spectral energy density of the unstable mode $k=2$ and its first seven overtones, scaled by the second mode angular frequency $\omega_0$, base speed of sound $a_0$, and rate of spectral energy transfer  $\mathcal{E}$~\eqref{eq:dissipation_rate}. (--), $T_H=450~\mathrm{K}$; (- -), $T_H=500~\mathrm{K}$; ($\cdot \cdot \cdot$), $T_H = 550~\mathrm{K}$.}
\label{fig:intro_spectral}
\end{figure}
Based on a scale-by-scale analysis of the growth of spectral energy density (figure~\ref{fig:intro_spectral}), the aforementioned three regimes of thermoacoustic wave amplification are identified as:
\begin{enumerate}
 \item \textbf{Modal growth}: Only the thermoacoustically unstable mode amplifies exponentially and all the other modes excited by the initial perturbation field decay (see \S~\ref{sec: Linear}). Higher harmonics begin to grow after $\alpha t \approx 3.2$ (figure~\ref{fig:intro_spectral}), setting the end of a purely harmonic growth. In the modal growth regime, the system is well approximated by the linear system of equations.
 \item \textbf{Hierarchical spectral broadening}: Energy cascades down to higher harmonics hierarchically: the $k^{\mathrm{th}}$ harmonic ($k>2$) grows at a rate equal to $k/2$-times the modal growth rate of the second harmonic (see \S~\ref{sec: NonlinearRegime_modelling} and \S~\ref{sec: nonlinear_cascade}), $\alpha$, that is
 \begin{equation}
 \alpha_k = \alpha k/2, \quad\, k \in \{4,6,8, \dots \}.
 \end{equation}
 The saturation of the spectral energy density $E_k$ occurs at $\alpha t \approx 10$  followed by the formation of resonating shock waves at limit cycle.
 \item \textbf{Shock dominated limit cycle}: In this regime ($\alpha t>10$), the continued injection of energy in the second-mode harmonic is balanced by the cascade of the spectral energy density into the overtones of the second mode and terminates by viscous dissipation at very high overtones ($k\approx 300$). The maximum number of overtones generated is a function of the acoustic phasing of the unstable mode. Moreover, at the limit cycle, the spectral energy density scales with instability growth rate $\alpha$ approximately as $\alpha^3$ (\S~\ref{sec: Spectral energy distribution}).
\end{enumerate}


\section{Harmonic growth analysis} \label{sec: Linear}

In this section, the amplification of acoustic waves in the linear regime in the proposed minimal-unit model is discussed. A system-wide differential eigenvalue problem is formulated and solved numerically utilizing the strategy adopted by~\cite{LinSH_JFM_2016} in (\S~\ref{sec:planar_thermoviscous_equations}). Utilizing the eigenvalue analysis, the computational setup has been optimized (\S~\ref{sec: optimization}) and the acoustic energy budgets are derived (\S~\ref{sec:Budgets}). This  provides analytical expressions for the cycle-averaged thermoacoustic production and dissipation, and elucidates the role of the acoustic phasing on thermoacoustic instability. Finally, the effects of varying the  hot-side temperature and the geometry on the thermoacoustic growth rates are discussed in~\S~\ref{sec: GrowthrateEffect}. While the eigenvalue analysis is restricted to the minimal-unit model, the physical conclusions and analysis presented in this section hold for any thermoacoustically unstable device operating in the low-acoustic amplitude regime.

\subsection{Linear thermoviscous quasi-planar wave equations} \label{sec:planar_thermoviscous_equations} \label{sec: LinearRegime}

The time-domain linear thermoviscous governing equations for a two-dimensional perturbation are:
\begin{subequations}
\begin{eqnarray}
\frac{\partial \rho'}{\partial t} + \rho_0 \frac{\partial u'}{\partial x} + u'\frac{d\rho_0}{dx} + \rho_0\frac{\partial v'}{\partial y}=0,\label{eq: 1_mass}\\
\frac{\partial u'}{\partial t} + \frac{1}{\rho_0}\frac{\partial p'}{\partial x} - \nu_0 \frac{\partial^2 u'}{\partial y^2}=0,\label{eq: 1_mom}\\
\frac{\partial s'}{\partial t} + u'\frac{ds_0}{dx} - \frac{k_0}{\rho_0T_0}\frac{\partial^2 T'}{\partial y^2}= 0, \label{eq: 1_ent}
\end{eqnarray}
\label{eq: GoverningEqsLinear}
\end{subequations}
where primed variables $\left(\cdot\right)'$ represent the fluctuations in the corresponding quantities and the subscript $0$ denotes the base state.
Axial diffusion terms and the $y$-momentum equation in~\eqref{eq: GoverningEqsLinear} have been neglected based on the scaling analysis reported in Appendix~\ref{sec: app1}. Combining the cross-sectionally averaged~\eqref{eq: 1_mass} and~\eqref{eq: 1_ent} and cross-sectionally averaging~\eqref{eq: 1_mom}, accounting for isothermal and no-slip boundary conditions, yields 
\begin{subequations}
\begin{gather}
 \frac{\partial p'}{\partial t} + \frac{\rho_0 a^2_0}{h}\frac{\partial U'}{\partial x} = \frac{\rho_0a^2_0}{h}\frac{q'}{\rho_0C_pT_0},\label{eq: thermoacoustic_time_a}\\
 \frac{1}{h}\frac{\partial U'}{\partial t} + \frac{1}{\rho_0}\frac{\partial p'}{\partial x} = \frac{1}{h}\frac{\tau'_w}{\rho_0},\label{eq: thermoacoustic_time_b}
\end{gather}
\label{eq: thermoacoustic_time}
\end{subequations}
respectively, where $h$ denotes the cross-sectional width of the duct, $U'$ denotes the fluctuations in the flow rate,
\begin{equation}
 U' = \int^{+h/2}_{-h/2}u'(x,y,t)dy,
\end{equation}
whereas $q'$ and $\tau'_w$ are the wall-heat flux and the wall-shear, respectively:
\begin{equation}
q' = 2k_0\left.\frac{\partial T'}{\partial y}\right|_{y=+h/2},\quad \tau'_w = 2\mu_0\left.\frac{\partial u'}{\partial y}\right|_{y=+h/2}.
\end{equation}
In sections $a, c,$ and $d$, linear wave propagation is assumed to be inviscid and adiabatic, hence isentropic, resulting in $q'=0$ and $\tau'_w=0$. Applying the normal mode assumption to~\eqref{eq: thermoacoustic_time}, namely,
\begin{equation}
 p'(x,t) = \hat{p}(x)e^{\sigma t},\quad U'(x,t) = \hat{U}(x)e^{\sigma t},
 \label{eq:normal_modes}
 \end{equation}
where $\sigma = \alpha + i\omega$ is the complex eigenvalue of the system with growth rate $\alpha$ and angular frequency $\omega$, leads to the thermoviscous set of quasi-planar wave equations in the frequency domain~\citep{LinSH_JFM_2016}:
\begin{subequations}
\begin{gather}
\sigma\hat{p} = \frac{\rho_0 a^2_0}{h_b}\left[\frac{1}{1+\left(\gamma -1\right)f_k}\left(\frac{\Theta(f_k - f_{\nu})}{(1-f_{\nu})(1-Pr)} - \frac{d}{dx}\right)\right]\hat{U},~\Theta = \frac{1}{T_0}\frac{dT_0}{dx},\label{eq: eig_p}\\
\sigma\hat{U} = -\frac{h_b}{\rho_0}\left(1-f_{\nu}\right)\frac{d\hat{p}}{dx},\label{eq: eig_u}
\end{gather}
\label{eq: thermoacoustic_eig}
\end{subequations}
where the thermoviscous functions $f_\nu$ and $f_k$ are given by
\begin{equation}
 f_{\nu} = \frac{\tanh(\eta h_b/2)}{\eta h_b/2},\quad f_k = \frac{\tanh(\eta h_b\sqrt{Pr}/2)}{\eta h_b\sqrt{Pr}/2},~~\mathrm{with}~~ \eta = \sqrt{i\omega/\nu_0}.
 \label{eq: thermoviscous}
\end{equation}

\subsection{Acoustic energy budgets for quasi-planar wave perturbations} \label{sec:Budgets}
Multiplying~\eqref{eq: thermoacoustic_time_a} by $p'/(\rho_0a^2_0)$ and~\eqref{eq: thermoacoustic_time_b} by $\rho_0 U'/h_b$ and adding them, yields the conservation equation
\begin{equation}
 \frac{\partial E}{\partial t} + \frac{\partial \mathcal{I}}{\partial x} = \mathcal{P}-\mathcal{D}, \label{eq: energy}
\end{equation}
for the one-dimensional acoustic energy density
\begin{equation} \label{eq:energy_planar}
 E = \frac{1}{2}\frac{p'^2}{\rho_0a^2_0} + \frac{1}{2}\rho_0\left(\frac{U'}{h_b}\right)^2,
 \end{equation}
consistent with the definition~\eqref{eq:energy_time}. The instantaneous acoustic flux $\mathcal{I}$ and the net energy production $\mathcal{P}-\mathcal{D}$ therein are given by
\begin{equation}
\mathcal{I} = \frac{p'U'}{h_b},\quad\mathcal{P}-\mathcal{D} = \frac{p'q'}{h_b\rho_0C_p T_0} + \frac{\tau'_w U'}{h_b^2}. 
\label{eq: energy_density_production}
\end{equation}
Averaging~\eqref{eq: energy} over one acoustic cycle and integrating axially over the periodic domain, $L$, yields:
\begin{equation}
\frac{d}{d(\epsilon t)}\int_L \overline{E}(x,\epsilon t) dx = \int_L \left(\overline{\mathcal{P}}-\overline{\mathcal{D}}\right) dx= \mathcal{R} ,
\label{eq: ModifiedRayleigh}
\end{equation}
where $\mathcal{R}$ is the Rayleigh index and $\epsilon t$ is the slow time scale (cf.~\eqref{eq:energy_spectral}). Relation \eqref{eq: ModifiedRayleigh} allows to unambiguously identify the onset of an instability via the criterion $\mathcal{R}>0$. This expression also accounts for wall-shear and wall-heat flux losses outside the regenerator (if present), which attenuate the thermoacoustic instability. Such thermoviscous losses are captured in the heat flux $q'$ and shear stress $\tau'_w$ terms (cf. \eqref{eq: energy_density_production}) in the respective duct sections. Utilizing the frequency domain linear equations~\eqref{eq: thermoacoustic_eig}, the wall-heat flux $\hat{q}$ and wall-shear $\hat{\tau}_w$ in the frequency domain are given by 
\begin{subequations}
\begin{gather}
 \hat{\tau}_w = h_b\frac{\partial \hat{p}}{\partial x}f_\nu,\\
 \hat{q} = h_b(i\omega)C_p T_0 \left(\frac{\Theta}{\left(1-Pr\right)\omega^2}\frac{\partial \hat{p}}{\partial x}\left(f_k - f_\nu\right) - \frac{\gamma - 1}{a^2_0}\hat{p}f_k\right).
\end{gather}
\label{eq: av_heat}
\end{subequations}
 \begin{figure}
 \centering
 \includegraphics[width=\textwidth]{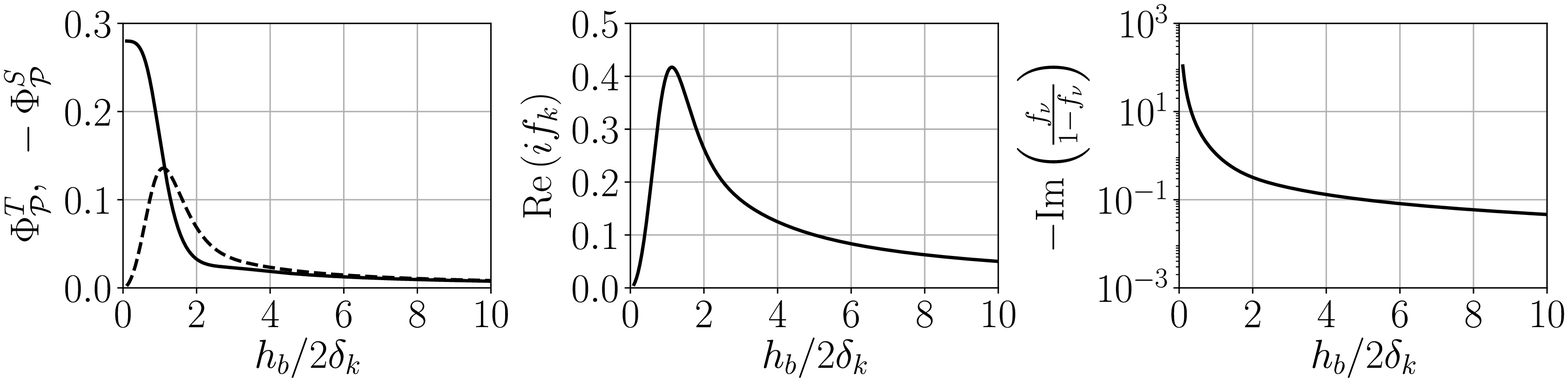}
\put(-385,5){$(a)$}
\put(-255,5){$(b)$}
\put(-130,5){$(c)$}
  \caption{Variation of thermoviscous functionals affecting the cycle averaged thermoacoustic production  $\overline{\mathcal{P}}$ ($a$), and dissipation $\overline{\mathcal{D}}$ ($b$ and $c$) of acoustic energy density versus the ratio of the regenerator half-width $h_b/2$ to the Stokes boundary layer thickness $\delta_k$ (cf.~\eqref{eq: energy_production}). ($a$): (--), $\Phi^T_{\mathcal{P}}$; (- -), $-\Phi^S_{\mathcal{P}}$.}
\label{fig: ProductionTerms}
\end{figure}
Combining~\eqref{eq: energy_density_production},~\eqref{eq: av_heat}, and~\eqref{eq: eig_u}, an analytical expression  is obtained for the cycle-averaged production $\overline{\mathcal{P}}$ and dissipation $\overline{\mathcal{D}}$ of the acoustic energy density~\eqref{eq:energy_planar}: 
\begin{subequations}
\begin{gather}
\overline{\mathcal{P}} = \frac{\Theta}{2\left(1-Pr\right)h_b}\left[\Phi^T_{\mathcal{P}}\;\mathrm{Re}\left(\hat{p}^*\hat{U}\right) -\Phi^S_{\mathcal{P}}\;\mathrm{Im}\left(\hat{p}^*\;\hat{U}\right)\right], \label{eq: budgets_production}\\ 
\overline{\mathcal{D}} = \mathrm{Re}\left(if_k\right) \frac{\omega\left(\gamma -1\right)}{2\rho_0a^2_0}|\hat{p}|^2 -\mathrm{Im}\left(\frac{f_\nu}{1-f_\nu}\right)\,\frac{\rho_0\omega}{2h^2_b}|\hat{U}|^2.\label{eq: budgets_dissipation}
\end{gather}
\label{eq: energy_production}
\end{subequations}
In the above relations, $\Phi^{T}_\mathcal{P}$ and $-\Phi^{S}_\mathcal{P}$ weigh the contributions to the thermoacoustic energy production by the travelling-wave, $\mathrm{Re}(\hat{p}^*\hat{U})$, and the standing-wave, $\mathrm{Im}(\hat{p}^*\,\hat{U})$, components respectively. Their expressions read:
\begin{equation} \label{eq:travelling_standing_weights}
\Phi^T_{\mathcal{P}} = \mathrm{Re}\left(\frac{f_k - f_\nu}{1-f_\nu}\right), \quad 
\Phi^S_{\mathcal{P}} = \mathrm{Im}\left(\frac{f_k - f_\nu}{1-f_\nu}\right).
\end{equation}
\begin{figure}
\includegraphics[width=0.75\textwidth]{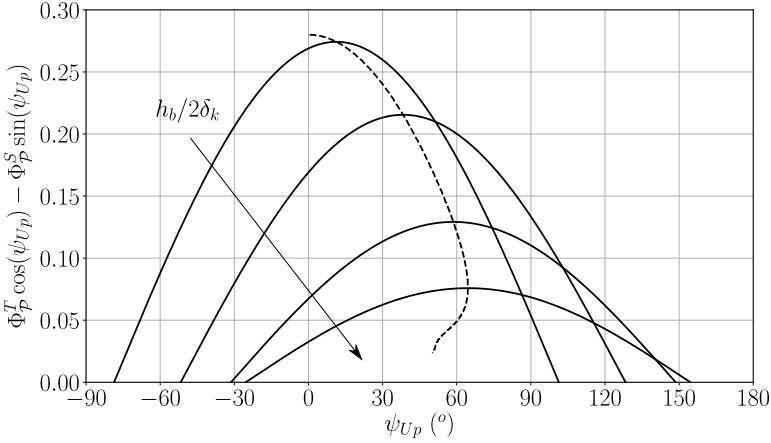}
\caption{Normalized thermoacoustic production $\Phi^T_{\mathcal{P}}\cos(\psi_{Up}) - \Phi^S_{\mathcal{P}}\sin(\psi_{Up})$ ~(cf.~\eqref{eq: budgets_production}) versus the phase angle difference between $\hat{p}$ and $\hat{U}$, $\psi_{Up}=\angle{\hat{U}}-\angle{\hat{p}}$, for increasing values of $h_b/2\delta_{k}=~$0.5, 1.0, 1.5, 2.0. (--); (- -), Optimum phasing maximizing the thermoacoustic production for continually varying $h_b/2\delta_{k}$.}
\label{fig: OptPhase}
\end{figure}
For $h_b/2\delta_k\leq1.13$, $|\Phi^T_{\mathcal{P}}|>|\Phi^S_{\mathcal{P}}|$, which implies that the regenerator half-width $h_b$ must remain comparable to or smaller than the Stokes boundary layer thickness $\delta_k$ to achieve higher thermoacoustic amplification of travelling waves ($\psi_{Up}\simeq 0^{\circ}$) (figures~\ref{fig: ProductionTerms}$a$ and~\ref{fig: OptPhase}). However, to maximize thermoacoustic energy production for standing waves ($\psi_{Up} \simeq \pm 90^{\circ}$), a larger regenerator half-width ($h_b/2\delta_k > 1.13$) is required. While production alone for a purely travelling wave ($\psi_{Up}=0$) is maximized in the limit $h_b/2\delta_k\rightarrow0$ (figure~\ref{fig: OptPhase}), for fixed temperature settings, dissipation also diverges (figure~\ref{fig: ProductionTerms}$c$). Therefore, \emph{pure travelling wave phasing, if at all achieved, always results in smaller net production of acoustic energy density compared to an optimal combination of standing and travelling waves}. Moreover, varying the temperature inside the regenerator results in a local variation of the ratio $h_b/2\delta_k$ which, in turn,  causes the optimum phase to vary along the regenerator (figure~\ref{fig: OptPhase}). For the temperature settings considered here (table~\ref{tbl:base_state}), the optimum phasing angle, averaged over the regenerator length changes from $43.17^{\circ}$ to $36.41^{\circ}$ as the temperature $T_{H}$ is increased.

\subsection{Effects of temperature gradient and geometry on growth rates}
\label{sec: GrowthrateEffect}
\begin{figure}
 \centering
\includegraphics[width=0.8\textwidth]{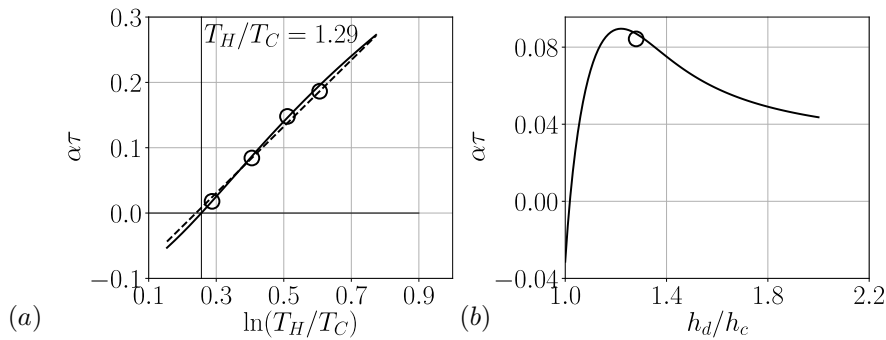}
\put(-330,5){$(a)$}
\put(-160,5){$(b)$}
  \caption{Dimensionless thermoacoustic growth rate, $\alpha\tau$ (cf.~\eqref{eq:tau}), versus the natural logarithm of the temperature ratio $T_H/T_C$ for $T_C=300~$\,K ($a$) and resonator area ratio $h_d/h_c$ ($b$). ($a$): (--), Linear stability analysis; (- -), logarithmic estimate of $\alpha\tau$~\eqref{eq: functional_approx} fitted using values at $T_H = 450\,\mathrm{K}$ and $T_H = 550\,\mathrm{K}$; ($\circ$), Navier-Stokes simulations.}
\label{fig: GrowthRate}
\end{figure}
The variation of the instability growth rates with the temperature and the geometry is further analysed utilizing the acoustic energy budget formulation developed in the previous section. To this end, the production, dissipation, and Rayleigh index (normalized by the pressure amplitude) are plotted in the convenient dimensionless forms 
\begin{equation}
 \overline{\mathcal{P}}_* = \frac{\overline{\mathcal{P}}\tau}{\rho_0a^2_0},\quad \overline{\mathcal{D}}_{*} = \frac{\overline{\mathcal{D}}\tau}{\rho_0a^2_0},\quad\mathcal{R}_* = \frac{\mathcal{R}\tau}{h_b\rho_0a^2_0}.\label{eq: dimensionless_PDR}
 \end{equation}
Cycle averaged production of acoustic energy density due to travelling wave and standing wave components given by
 \begin{subequations}
\begin{eqnarray}
 \mathcal{R}_{T*} = \frac{\tau}{2(1-Pr)\rho_0a^2_0h^2_b}\int_b\Theta\,\mathrm{Re}\left(\hat{p}^*\hat{U}\right)\Phi^T_{\mathcal{P}}dx ,\label{eq: dimensionless_RT}\\ \mathcal{R}_{S*} = -\frac{\tau}{2(1-Pr)\rho_0a^2_0h^2_b}\int_b\Theta\,\mathrm{Im}\left(\hat{p}^*\hat{U}\right)\Phi^S_{\mathcal{P}}dx, \label{eq: dimensionaless_RS}
\end{eqnarray}
\end{subequations}
respectively, are also analysed, where
\begin{equation} \label{eq:tau}
\tau = h^2_b/\nu_0
\end{equation}
is a reference viscous time scale in the regenerator with $\nu_0$ evaluated at $T_C=300\,$K.

\begin{figure}
\centering
\includegraphics[width=0.8\textwidth]{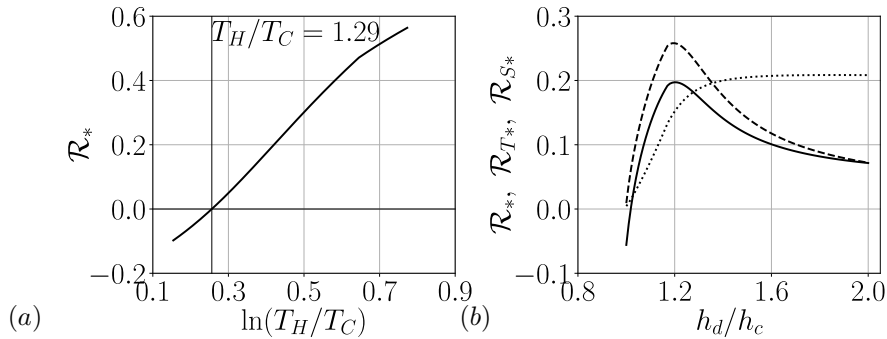}
\put(-330,5){$(a)$}
\put(-160,5){$(b)$}
\caption{Dimensionless Rayleigh index $\mathcal{R}_*$ versus the natural logarithm of the temperature ratio $T_H/T_C$ for $T_C=300\,$K ($a$) and the resonator area ratio $h_d/h_c$ ($b$). ($b$): (--), $\mathcal{R}_*$ ; (- -), $\mathcal{R}_{T*}$ ; ($\cdot \cdot \cdot$), $\mathcal{R}_{S*}$.}
\label{fig: RayleighIndex}
\end{figure}

Increasing the hot side temperature $T_H$, the thermoacoustic production $\overline{\mathcal{P}}_*$ increases monotonically, approximately as
\begin{equation}
 \overline{\mathcal{P}}_* \sim \Theta = \frac{d}{dx}\ln[T_0(x)],
 \label{eq: ApproxProdDiss}
\end{equation}
and more rapidly than the dissipation $\overline{\mathcal{D}}_*$, yielding positive values of the Rayleigh index $\mathcal{R}_*$ (figure~\ref{fig: RayleighIndex}$a$) for $T_H/T_C>1.29$ (figure~\ref{fig: GrowthRate}$a$). The Rayleigh index can thus be used to quantify the thermoacoustic growth rate (figure~\ref{fig: GrowthRate}$a$,~\ref{fig: RayleighIndex}$a$) as
\begin{equation}
 \mathcal{R}_*\sim \alpha\tau \approx A \ln(T_H/T_C) - B,
 \label{eq: functional_approx}
\end{equation}
where $A$ and $B$ are geometry dependent fitting coefficients. 
With increasing resonator area ratio $h_d/h_c$, keeping $h_c$ fixed, the frequency decreases monotonically by approximately $4\%$ in the range of $h_d/h_c$ considered. However, the growth rates vary non-monotonically, reaching a local maximum  at $h_d/h_c \approx 1.28$ (figure~\ref{fig: GrowthRate}$b$). 
Also, the net production of acoustic energy density due to the travelling wave component $\mathcal{R}_{T*}$ (figure~\ref{fig: RayleighIndex}$b$) peaks at $h_d/h_c\approx1.28$. Moreover, the high mechanical impedance of the section $d$ ($\rho_0a_0h_d$) for high values of $h_d$ increases the standing wave component of the acoustic power $\mathcal{R}_{S*}$ and decreases the travelling wave  component $\mathcal{R}_{T*}$. The variation of the transverse geometrical parameters does not significantly alter the frequency. Therefore, changes in the ratio $(f_k - f_\nu)/(1-f_{\nu})$ are also negligible and do not influence the growth rates significantly (cf.~\eqref{eq:travelling_standing_weights}).

\section{Formulation of a Nonlinear Thermoacoustic Model}
\label{sec: NonlinearRegime_modelling}
As a result of the modal thermoacoustic instability, large pressure amplitudes ($\sim 160~\mathrm{dB}$) are generated, which result in the nonlinear steepening of the waveform. In the spectral space, the nonlinear steepening can be viewed as the cascade of energy from the unstable mode into higher harmonics with correspondingly shorter wavelengths. Moreover, inside the regenerator, large amplitude perturbations in thermodynamic quantities are responsible for thermoacoustic nonlinearities. As a result of nonlinear wave propagation, thermoacoustically sustained shock waves propagate in the system. While the analysis above highlights that quasi-travelling wave phasing is essential for high thermoacoustic growth rates, nonlinear steepening is also favoured by such phasing~\citep{BiwaEtAl_JASA_2014}.

In this section, a first-principles-based theoretical framework accounting for acoustic and thermoacoustic nonlinearities up to second order is developed, and a quasi one-dimensional evolution equation is obtained for nonlinear thermoacoustic waves~\eqref{eq: nonlinear_P}. In \S~\ref{sec: governing_equations} mass, momentum, and energy (combined with the second law of thermodynamics) equations correct up to second order are introduced. Furthermore, in \S~\ref{sec: nonlinear_wave} and \S~\ref{sec: nonlinear_thermoacoustic}, cross-sectionally averaged nonlinear spatio-temporal evolution model equations are derived with the time-domain approximations of wall-shear and wall-heat flux outlined in~\S~\ref{sec: shear_heat}. 
\subsection{Governing equations for nonlinear thermoviscous perturbations}
\label{sec: governing_equations}
The nonlinear governing equations, correct up to second order, for a two-dimensional perturbation read:
\begin{subequations}
\begin{gather}
\frac{\partial \rho'}{\partial t} + \rho_0 \frac{\partial u'}{\partial x} + u'\frac{d\rho_0}{dx} + \rho_0\frac{\partial v'}{\partial y}= \left[-\rho'\frac{\partial u'}{\partial x} - u'\frac{\partial \rho'}{\partial x}\right],\label{eq: 2_mass}\\
\frac{\partial u'}{\partial t} + \frac{1}{\rho_0}\frac{\partial p'}{\partial x} - \nu_0 \frac{\partial^2 u'}{\partial y^2} - \frac{1}{\rho_0}\frac{\partial }{\partial x}\left[\mu_0\left(\xi_B + \frac{4}{3}\right)\frac{\partial u'}{\partial x}\right]=  \left[-\frac{\rho'}{\rho_0}\frac{\partial u'}{\partial t} - \frac{1}{2}\frac{\partial u'^2}{\partial x}\right],\label{eq: 2_mom}\\
\frac{\partial s'}{\partial t} + u'\frac{ds_0}{dx} - \frac{Rk_0}{p_0}\frac{\partial^2 T'}{\partial y^2} - \frac{R}{p_0}\frac{\partial}{\partial x}\left(k_0\frac{\partial T'}{\partial x}\right)= \left[-\frac{p'}{p_0}\left(\frac{\partial s'}{\partial t} + u'\frac{ds_0}{dx}\right) - u'\frac{\partial s'}{\partial x}\right], \label{eq: 2_ent}
\end{gather}
\end{subequations}
where, again, primed variables $\left(\cdot\right)'$ represent the fluctuations in corresponding quantities, the subscript $0$ denotes the base state, whereas $\xi_B=2/3$ is the ratio of the bulk viscosity coefficient $\mu_B$ to the shear viscosity coefficient $\mu$. The terms on the left-hand side are linear in the perturbation variables while those on the right-hand side are nonlinear. The entropy generation due to viscous dissipation is neglected, as well are the pressure gradients and velocity in $y$ direction (boundary layer assumption), and the fluctuations in the  diffusivity coefficients $\mu$ and $k$ (see appendix~\ref{sec: app1}). Higher harmonics with correspondingly shorter wavelengths are generated due to the nonlinear spectral energy cascade. Consequently, the axial diffusion terms in~\eqref{eq: 2_mom} and~\eqref{eq: 2_ent}, which have been neglected in~\S~\ref{sec: Linear}, become significant and act as the primary sink of energy at large harmonic scales in the spectral space. 

We seek to collapse~\eqref{eq: 2_mass}--\eqref{eq: 2_ent} to obtain a set of equations similar to~\eqref{eq: AcousticDynamicalSystem}. To this end, the following quadratic thermodynamic constitutive equation relating the density fluctuations $\rho'$ with the pressure and entropy fluctuations ($p'$ and $s'$, respectively) is considered: 
\begin{equation}
\rho' = \alpha_s p' + \alpha_ps' + \frac{1}{2}\left(\beta_s p'^2 + \beta_p s'^2 + 2\beta_{sp} s'p'\right), \label{eq: constitutive_eqn}
\end{equation}
where the thermodynamic coefficients $\alpha $ and $\beta$ are given by
\begin{subequations}
\begin{gather}
 \alpha_s = \left(\frac{\partial \rho}{\partial p}\right)_s = \frac{1}{a^2_0},\quad \alpha_p = \left(\frac{\partial \rho}{\partial s}\right)_p = -\frac{\rho_0}{C_p},\\
\beta_s = \left(\frac{\partial^2 \rho}{\partial p^2}\right)_s = -\frac{\gamma -1}{\rho_0 a^4_0},\quad \beta_p = \left(\frac{\partial^2 \rho}{\partial s^2}\right)_p = \frac{\rho_0}{C^2_p},\\
 \beta_{sp} = \left[\frac{\partial }{\partial s}\left(\frac{\partial \rho}{\partial p}\right)_s\right]_p = \left[\frac{\partial }{\partial s}\left(\frac{\rho}{\gamma p}\right)\right]_p = -\frac{1}{C_p a^2_0}.\label{eq: const_coeff}
\end{gather}
\end{subequations}
\begin{figure}
\includegraphics[width=0.7\textwidth]{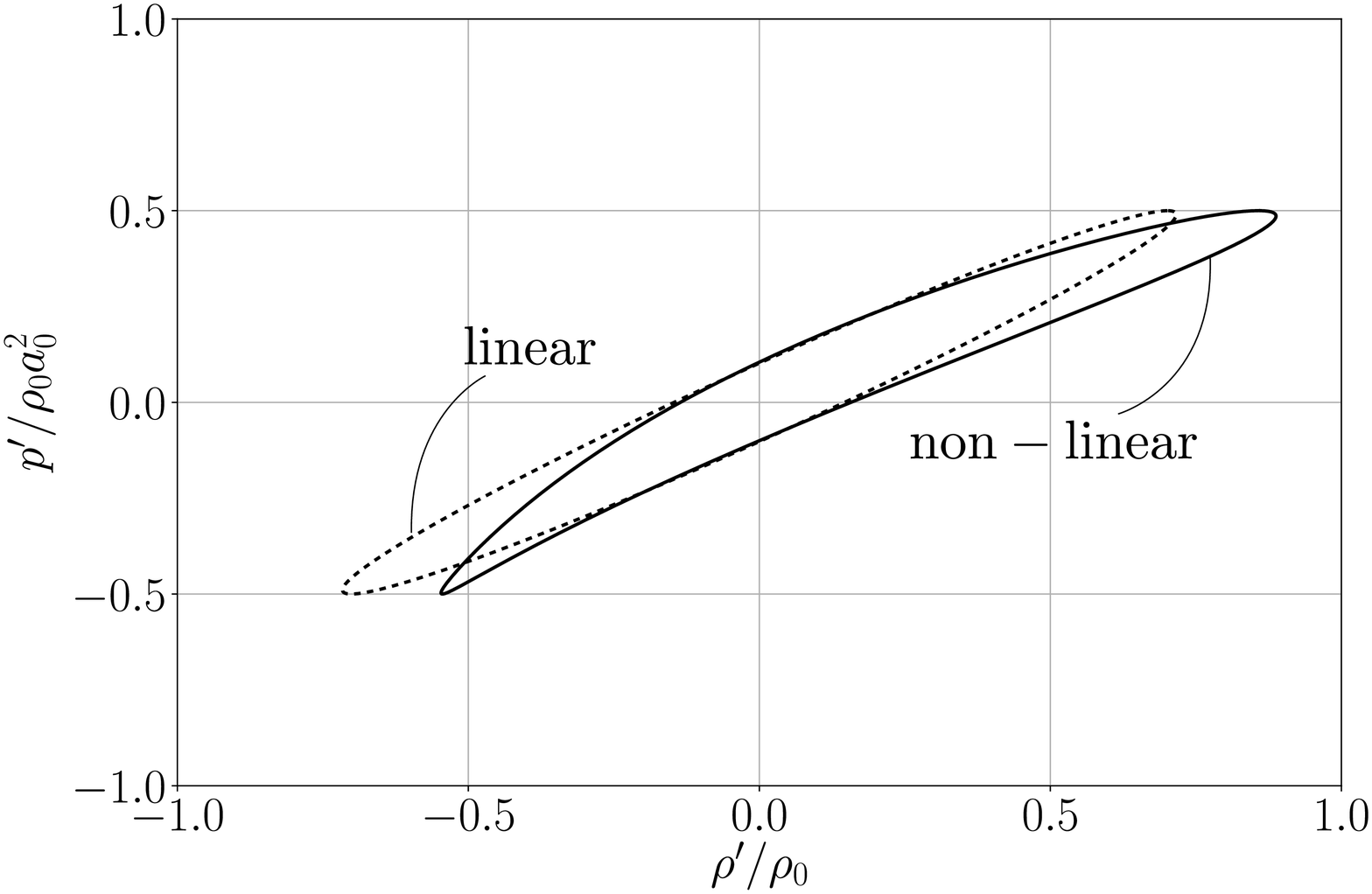}
\caption{Thermodynamic cycle in $p'-\rho'$ plane for non-isentropic thermoacoustic wave amplification. (--), Nonlinear; (- -), Linear. Sample perturbation fields are: $p' = 0.5p_0\cos\omega t$ and $s' = -Rp'/p_0 + 250\sin\left(\omega(t + \tau)\right)$ for $\tau = h^2_b/\nu = 0.0107~\mathrm{s}$.}
\label{fig: constitutive}
\end{figure}
The coefficients $\alpha_p$ and $\alpha_s$ contribute to first order wave propagation and thermoacoustic effects while the second-order coefficients $\beta_s,~\beta_p,$ and $\beta_{sp}$ in~\eqref{eq: constitutive_eqn} account for the corresponding nonlinear effects.~\cite{HedbergR_JAP_2011} have demonstrated the hysteretic effects of nonlinear wave propagation retaining only the $\beta_s$ nonlinear term and $\alpha_p$ term accounting for irreversible entropy changes. Figure~\ref{fig: constitutive} shows the hysteresis cycle for the second order constitutive relation in~\eqref{eq: constitutive_eqn} in comparison with the first order approximation.
Nonlinear wave propagation outside the regenerator (sections $a$, $c$, and $d$) is not affected by the no-slip and isothermal boundary conditions. Only higher order irreversible entropy fluctuations are generated due to the axial conduction terms~\citep{Hamilton_NLA_1998}. Consequently, the constitutive relation~\eqref{eq: constitutive_eqn} can be approximated with the terms corresponding to $\alpha_s, \alpha_p$, and $\beta_s$ retained. However, inside the regenerator, wall-shear and wall-heat flux from the no-slip isothermal boundaries generate first order reversible entropy fluctuations; hence, in order to capture the nonlinear thermoacoustic wave amplification inside the regenerator, nonlinear terms in entropy perturbations need to be included, as shown in~\eqref{eq: constitutive_eqn}. Starting from the general second order governing equations discussed above, the spatio-temporal evolution equations for the fluctuations in pressure $p'$ and flow rate $U'$ are derived in the following sections.

\subsection{Free-shear nonlinear wave propagation}
\label{sec: nonlinear_wave}
Waves outside the regenerator (sections $a$, $c$, and $d$) in the minimal unit setup (figure~\ref{fig:computational_setup}) propagate in the absence of wall-shear and wall-heat flux. As a result, terms involving the transverse gradient of $u'$ in~\eqref{eq: 2_mom} and $T'$ in~\eqref{eq: 2_ent} can be neglected and entropy fluctuations remain second order in the nonlinear regime as well~\citep{Hamilton_NLA_1998}. Up to second order, the wave propagation in the duct sections $a$, $c$, and $d$ is governed by
\begin{gather}
   \frac{\partial p'}{\partial t} = -\frac{\gamma p_0}{h}\left(1 + \frac{1+\gamma}{\gamma}\frac{p'}{p_0}\right)\frac{\partial U'}{\partial x} + \frac{k}{\rho_0}\left(\frac{1}{C_v} - \frac{1}{C_p}\right)\frac{\partial^2 p'}{\partial x^2},\label{eq: inviscid_p}\\
   \frac{\partial U'}{\partial t} = -\frac{h}{\rho_0}\frac{\partial p'}{\partial x} + \nu_0\left(\xi_B + \frac{4}{3}\right)\frac{\partial^2 U'}{\partial x^2},\label{eq: inviscid_U}
  \end{gather}
where $h = h_a,~h_c,$ or $h_d$. Second-order nonlinearities in~\eqref{eq: inviscid_p} cause the waveform distortion and steepening.

\subsection{Wall-shear and wall-heat flux}
\label{sec: shear_heat}
By design, the regenerator width is comparable to the local viscous and thermal Stokes layer thickness ($h_b/2\delta_\nu\sim1$, $h_b/2\delta_k\sim1$). Due to the wall-shear, the velocity fluctuations inside the regenerator vary in the $y$ direction as well. Moreover, the wall-heat flux contributes to the first order entropy fluctuations. Noting that nonlinear acoustic waves can be decomposed into acoustic, viscous, and entropic modes~\citep{ChuK_JFM_1958, Pierce_1989}, the following decomposition for the entropy and velocity fluctuations inside the regenerator is considered:
\begin{equation}
 u'(x,y,t) = \tilde{u}(x,t) + u'_{\nu}(x,y,t), \quad s' = \tilde{s}(x,t)+ s'_{q}(x,y,t),
\end{equation}
where $u'_{\nu}$ is the viscous velocity fluctuation and $\tilde{u}$ is the nonlinear acoustic wave field. The former is diffused by viscosity and is governed by the unsteady diffusion equation 
\begin{equation}
 \frac{\partial u'_{\nu}}{\partial t} = \nu_0\frac{\partial^2 u'_{\nu}}{\partial y^2},\quad u'_{\nu}(x,y=\pm h_b/2,t) = -\tilde{u}(x,t). \label{eq: shear}
\end{equation}
Similarly, $\tilde{s}$ accounts for the entropy changes due to the nonlinear acoustic wave propagation and $s'_q$ corresponds to the first order entropy changes due to the wall-heat flux inside the regenerator and is governed by the equation
\begin{equation}
 \frac{\partial s'_q}{\partial t} + u'\frac{ds_0}{dx} = \frac{\nu_0}{Pr}\frac{\partial^2 s'_q}{\partial y^2},\quad s'_q\left(x,y=\pm h_b/2, t\right) = -\tilde{s} + s'_\mathrm{w}, \label{eq: heat_transfer}
\end{equation}
where $s'_\mathrm{w}$ corresponds to the entropy fluctuations at the isothermal walls driven by pressure fluctuations
\begin{equation}
 s'_\mathrm{w} = -\frac{1}{\rho_0 T_0}p^{\prime} = -\frac{R}{p_0}p^{\prime}. \label{eq: entropy_wall}
\end{equation}
Equations~\eqref{eq: shear} and~\eqref{eq: heat_transfer} suggest the following infinite series solution forms for the viscous and the entropic fields: 
\begin{subequations}
\begin{gather}
 u'_{\nu} = -\tilde{u} + \sum_{j=0}^{\infty}\check{u}_j(x,t) \cos(\zeta_j y),\quad s'_q = -\tilde{s} + s_\mathrm{w} + \sum_{j=0}^{\infty}\check{s}_j(x,t)\cos \left(\zeta_j y\right),\label{eq: eigenfunctions_y}\\
\quad\text{with}\quad \zeta_j = \left(2j+1\right)\frac{\pi}{h_b}.\label{eq: wavenumber_y}
\end{gather}
\end{subequations}
Performing eigenfunction expansions along the $y$ direction~\eqref{eq: eigenfunctions_y} yields the following evolution equations for the Fourier coefficients corresponding to the viscous and entropic modes:
\begin{eqnarray}
 \frac{\partial \check{u}_j}{\partial t} + \nu_0 \zeta^2_j \check{u}_j = (-1)^{j+1}\frac{2}{\zeta_j h_b}\left(\frac{1}{\rho_0}\frac{\partial p'}{\partial x}\right),\label{eq: u_modes}\\
 \frac{\partial \check{s}_j}{\partial t} + \check{u}_j\frac{d s_0}{dx} + \frac{\nu_0}{Pr}\zeta^2_j \check{s}_j = (-1)^{j}\frac{2 R}{\zeta_jh_b p_0}\frac{\partial p'}{\partial t}.
 \label{eq: s_modes}
\end{eqnarray}
Equations~\eqref{eq: u_modes} and~\eqref{eq: s_modes} determine the evolution of the transverse modes of the longitudinal velocity $\check{u}_j$ and entropy $\check{s}_j$ fluctuations. In Appendix~\ref{sec: app2}, the convergence properties of the infinite series in~\eqref{eq: eigenfunctions_y} for $j\rightarrow\infty$ are discussed.

\subsection{Nonlinear thermoviscous wave equations}
\label{sec: nonlinear_thermoacoustic}

The axial velocity fluctuations $u'$ are governed by~\eqref{eq: 2_mom} up to second order accuracy. However, nonlinearities in~\eqref{eq: 2_mom} result in acoustic streaming, which is suppressed in the current analysis, and is therefore neglected~\citep{Hamilton_NLA_1998}. Integrating the resulting momentum equation in $y$ and substituting~\eqref{eq: eigenfunctions_y} yields 
\begin{eqnarray}
\frac{\partial U'}{\partial t} + \frac{h_b}{\rho_0}\frac{\partial p'}{\partial x} =  \tau'_w + \frac{1}{\rho_0}\frac{\partial }{\partial x}\left[\mu_0\left(\xi_B + \frac{4}{3}\right)\frac{\partial U'}{\partial x}\right],\label{eq: nonlinear_U}
\end{eqnarray}
where
\begin{eqnarray}
\tau'_w =  2\nu_0\sum^{\infty}_{j=0}(-1)^{j+1}\check{u}_j(x,t)\zeta_j.\label{eq: shear_stress}
\end{eqnarray}
Equation~\eqref{eq: inviscid_p} governs the evolution of the pressure fluctuations up to second order in the free-shear/adiabatic ducts. In order to derive an analogous governing equation for the regenerator,~\eqref{eq: 2_mass},~\eqref{eq: 2_ent}, and~\eqref{eq: constitutive_eqn} are combined to obtain 
\begin{eqnarray}
\underbrace{\frac{\partial p'}{\partial t} + \frac{\rho_0a^2_0}{h_b}\frac{\partial U'}{\partial x}}_\text{wave propagation}=\frac{\rho_0a^2_0}{h_b}\Bigg(\frac{q'}{C_p\rho_0T_0}+\underbrace{q_2  + \mathbb{T} -\mathbb{Q}}_{\substack{\text{thermodynamic}\\ \text{nonlinearities}}} +  \mathbb{D}_s\Bigg) - \mathbb{C}, \label{eq: nonlinear_P}
\end{eqnarray}
where 
\begin{equation}
 q' = \frac{2\nu_0\rho_0 T_0}{Pr}\sum^{\infty}_{j=0}(-1)^{j+1}\check{s}_j(x,t)\zeta_j,
 \label{eq: heat_flux}
\end{equation}
defines the fluctuating wall-heat flux and couples the pressure evolution~\eqref{eq: nonlinear_P} with the entropic mode evolution~\eqref{eq: s_modes}, whereas $\mathbb{Q}$ denotes the nonlinear interaction of pressure and wall-heat flux fluctuations, 
\begin{equation}
 \mathbb{Q} = \frac{\gamma p'q'}{C_p p_0\rho_0T_0},
 \label{eq: MacrosonicThermoacousticInteraction}
\end{equation}
 hereafter referred as \emph{macrosonic thermoacoustic interaction}. The term denoted by $q_2$ corresponds to the second order heat flux which is a quadratic function of the entropy gradient in $y$. The nonlinear terms denoted by the double faced $\mathbb{T}$ correspond to the constitutive (thermodynamic) nonlinearities which account for the second order density fluctuation due to first order entropic modes. The terms denoted by the double faced $\mathbb{C}$ correspond to the convective nonlinearities in~\eqref{eq: 2_mom} and~\eqref{eq: 2_ent} and those denoted by $\mathbb{D}_s$ account for the axial diffusion of gradients in highly nonlinear regimes of thermoacoustic wave amplification. A detailed derivation of~\eqref{eq: nonlinear_P} and expressions for the terms $q_2$, $\mathbb{D}_s$, $\mathbb{T}$, and $\mathbb{C}$ are given in Appendix~\ref{sec: app2}. Equations~\eqref{eq: u_modes},~\eqref{eq: s_modes},~\eqref{eq: nonlinear_U}, and~\eqref{eq: nonlinear_P} constitute the governing equations for the spatio-temporal evolution of large amplitude acoustic perturbations inside the regenerator. The macrosonic thermoacoustic interaction~\eqref{eq: MacrosonicThermoacousticInteraction} breaks the thermodynamic symmetry between the interactions of compressions and dilatations with the wall-heat flux inside the regenerator, thus highlighting that the entropy of a Lagrangian parcel of fluid changes by a small amount under high amplitude compressions ($\rho'>0$), compared to dilatations ($\rho'<0$), for the same amount of heat input or output. 

In general, for thermoacoustic devices in looped configuration, the length of the regenerator is very short compared to the total length of the device. As a result, higher order terms affecting only the propagation of the acoustic perturbations, such as convective nonlinearities, can be neglected inside the regenerator. Under such hypotheses,  the following approximate nonlinear governing equation for the pressure fluctuations $p'$ inside a short regenerator is obtained:
\begin{equation}
 \frac{\partial p'}{\partial t} + \frac{\rho_0a^2_0}{h_b}\frac{\partial U'}{\partial x}\approx \frac{\rho_0a^2_0}{h_b}\left\{\frac{1}{C_p}\left[\left(1-\frac{\gamma p'}{p_0}\right)\frac{q'}{\rho_0T_0}\right]\right\}. \label{eq: nonlinear_P_reduced}
\end{equation}
In the above equation, terms $\mathbb{T}$, $\mathbb{D}_s$, and $q_2$ are neglected for simplicity. 
Equations~\eqref{eq: u_modes} and~\eqref{eq: s_modes} can also be integrated in time analytically to express the wall-shear $\tau'_w$ and wall-heat flux $q'$ in terms of acoustic variables. 
The time integration of~\eqref{eq: u_modes} and~\eqref{eq: s_modes} yields: 
\begin{gather}
 \check{u}_j = (-1)^{j+1}\frac{2}{\zeta_j h_b\rho_0}\int^{t}_{-\infty}e^{-\frac{t - \eta}{\tau_j}}\frac{\partial p'}{\partial x}(x,\eta)d\eta, \label{eq: analytical_un}\\
 \check{s}_j = -\frac{ds_0}{dx}\int^{t}_{-\infty}e^{-\frac{t - \eta}{Pr\tau_j}}\check{u}_j(x,\eta)d\eta + (-1)^j\frac{2R}{\zeta_jh p_0}\int^{t}_{-\infty}e^{-\frac{t - \eta}{Pr\tau_j}}\frac{\partial p'}{\partial \eta}(x,\eta) d\eta. \label{eq: analytical_sn}
 \end{gather}
where $\tau_j = 1/\nu_0\zeta^2_j$ defines the viscous relaxation time for the $j^{\mathrm{th}}$ viscous mode.
Hence, writing the relaxation functional for some function $\phi(x,t)$, namely,
\begin{equation}
 \mathcal{G}_j(\phi, \tau_j) = \int^{t}_{-\infty}e^{-\frac{t-\eta}{\tau_j}}\phi(x,\eta)d\eta, \label{eq: relaxation_functional} 
\end{equation}
and summing~\eqref{eq: analytical_un} and~\eqref{eq: analytical_sn} over $j$, the following expressions for the wall-shear and wall-heat flux are obtained:
\begin{gather}
 \tau'_w = \frac{4\nu_0}{\rho_0 h_b}\sum^{\infty}_{j=0}\mathcal{G}_j\left(\frac{\partial p'}{\partial x}, \tau_j\right),\label{eq: analytical_tau}\\
  q' = \frac{2\rho_0\nu_0T_0}{Pr}\sum^{\infty}_{j=0}\left[(-1)^{j}\frac{ds_0}{dx}\mathcal{G}_j\left(\zeta_j\check{u}_j, \tau_jPr\right)-\frac{2R}{h_bp_0}\mathcal{G}_j\left(\frac{\partial p'}{\partial t}, \tau_j Pr\right)\right]. \label{eq: analytical_q}
\end{gather}
Equations~\eqref{eq: analytical_tau} and~\eqref{eq: analytical_q} provide first order expressions for the wall-shear and the wall-heat flux as a function of a generic acoustic field near the walls and, together with~\eqref{eq: nonlinear_U} and~\eqref{eq: nonlinear_P_reduced}, complete the nonlinear wave propagation model equations. However, in the present work,~\eqref{eq: u_modes} and~\eqref{eq: s_modes} have been considered for time integration for simplicity.

\cite{Sugimoto_JFM_2010} systematically derived the functional form of $\mathcal{G}_j$ for extremely \emph{thick} and \emph{thin} diffusion layers using the linear acoustic field approximation. The functional $\mathcal{G}_j$ in equation~\eqref{eq: relaxation_functional} approximates the wall-shear and the wall-heat flux in terms of any acoustic field, linear or nonlinear, though within the restriction of linear decomposition of the field into viscous and entropic modes. Recently,~\cite{Sugimoto_JFM_2016} also developed a theoretical framework to elucidate high amplitude nonlinear wave propagation in a shear dominated duct with the restriction of very \emph{thick} diffusion layers and focusing on streaming. However, the wall-shear and wall-heat flux expressed in terms of the acoustic field variables by the relaxation functional $\mathcal{G}_j$ in~\eqref{eq: analytical_tau} and~\eqref{eq: analytical_tau} hold true irrespective of the relative thickness between the diffusion layers and the channel width. 

\begin{figure}
\centering{\includegraphics[width=0.6\textwidth]{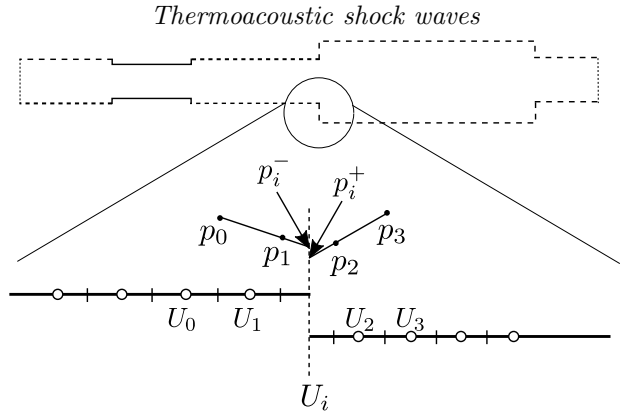}}
\caption{Solution technique for integrating quasi one-dimensional governing equations in time across abrupt area changes. Interface bulk velocity at $n$ time step, $U^{n}_i$ is calculated such that difference of left and right extrapolations of pressure ($p^{-}_i$ and $p^{+}_i$ respectively) equals minor loss $\Delta p_{\mathrm{ml}}$ at $n+1$ time step which is a function of $U^{n}_i$.}
\label{fig: solution_technique}
\end{figure}

\subsection{Solution technique, minor losses, and shock capturing}
\label{sec: solution_technique}
The model equations~\eqref{eq: inviscid_p},~\eqref{eq: inviscid_U},~\eqref{eq: u_modes},~\eqref{eq: s_modes},~\eqref{eq: nonlinear_U}, and~\eqref{eq: nonlinear_P} are integrated in time utilizing a $4^{\mathrm{th}}$ order explicit Runge-Kutta advancement with a second order staggered spatial discretization. Appropriate interface conditions for ducts with abrupt area jumps are provided as well. In general, abrupt area jumps cause minor losses in the pressure ($\Delta p_{\mathrm{ml}}$) due to nonlinear vortex generation which results in second order losses. Hence, minor losses can be modeled as quadratic functions of the first order interface flow rate,  i.e., $\Delta p_{\mathrm{ml}} \sim U^2_i$. In analogy with the approach followed by~\cite{LinSH_JFM_2016}, in order to match the difference between the left-hand pressure ($p^{-}_i$) and the right-hand pressure ($p^{+}_i$) limits at the interface to these minor losses, the linearly extrapolated pressures from the closest discretization points are used, as schematically shown in figure~\ref{fig: solution_technique}. Accordingly,  the interface condition imposed reads:
\begin{equation}
 U^{(2)}_i: p^{-}_i - p^{+}_i = \left(\frac{3}{2}p_1 - \frac{1}{2}p_0\right) - \left(\frac{3}{2}p_2 - \frac{1}{2}p_3\right) = \Delta p_{\mathrm{ml}} \approx \frac{1}{2}\rho_0K U^{(1)^2}_i,
 \label{eq: interface}
\end{equation}
where $K$ represents an empirically determined minor loss coefficient for an incompressible steady flow and $U^{(1)}_i$ corresponds to the first order interface flow rate calculated such that 
\begin{equation}
 U^{(1)}_i: p^{-}_i = p^{+}_i.
\end{equation}
Hence, the second order flow rate $U^{(2)}_i$ is calculated imposing the condition in~\eqref{eq: interface}.
In the limit of shock wave formation, the spectral energy cascade results in the formation of large gradients, which reach the limit of grid resolution. Local Legendre polynomial expansions of the pressure field are thus used to introduce an artificial bulk viscosity which increases the effective diffusion coefficient in the direction of propagation. Details of such a windowed artificial viscosity implementation are discussed in Appendix~\ref{sec: app3}.

\section{Nonlinear Spectral Energy Dynamics} 
\label{sec: nonlinear_cascade}

\begin{figure}
\includegraphics[width=1.0\textwidth]{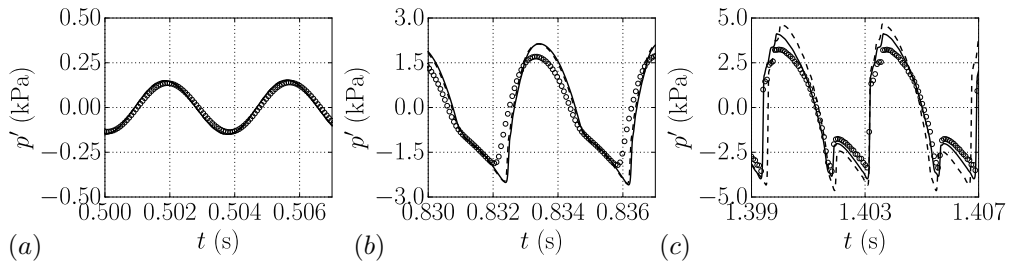}
\put(-380,5){$(a)$}
\put(-250,5){$(b)$}
\put(-135,5){$(c)$}
\caption{Comparison of filtered thermoacoustic signal as obtained from time integration of nonlinear model~\eqref{eq: inviscid_p},~\eqref{eq: inviscid_U},~\eqref{eq: u_modes},~\eqref{eq: s_modes},~\eqref{eq: nonlinear_U}, and~\eqref{eq: nonlinear_P} probed at $x=1.54~\mathrm{m}$ in harmonic growth regime ($a$), hierarchical spectral broadening regime ($b$), and limit cycle ($c$). (--), Model with macrosonic thermoacoustic interaction $\mathbb{Q}$; (- -), Model without $\mathbb{Q}$, ($\circ$), Navier-Stokes simulations.}
\label{fig: model_signal}
\end{figure}
In order to elucidate the physics of the hierarchical spectral broadening regime, the results from the nonlinear model derived in the previous section are discussed here and compared to the high-order Navier-Stokes calculations. Pressure time series from the time integration of~\eqref{eq: nonlinear_P_reduced} are in fairly good agreement with the fully compressible Navier-Stokes simulations (see figure~\ref{fig: model_signal}). Since several nonlinearities are neglected in calculations via~\eqref{eq: nonlinear_P_reduced}, time integration results in spurious temporal variations of the time averaged pressure fluctuations which are removed in further discussions. The time integration of the nonlinear model up to the limit cycle offers a significant reduction in computational cost (about $500$ times faster than the fully compressible Navier-Stokes simulations) and predicts the captured limit cycle amplitudes within $80\%$ accuracy. Additionally, upon excluding the macrosonic thermoacoustic interaction term from the model~\eqref{eq: MacrosonicThermoacousticInteraction}, the accuracy of the predicted limit cycle amplitude gets reduced to $60\%$. 

\begin{figure}
\centering{\includegraphics[width=1.0\textwidth]{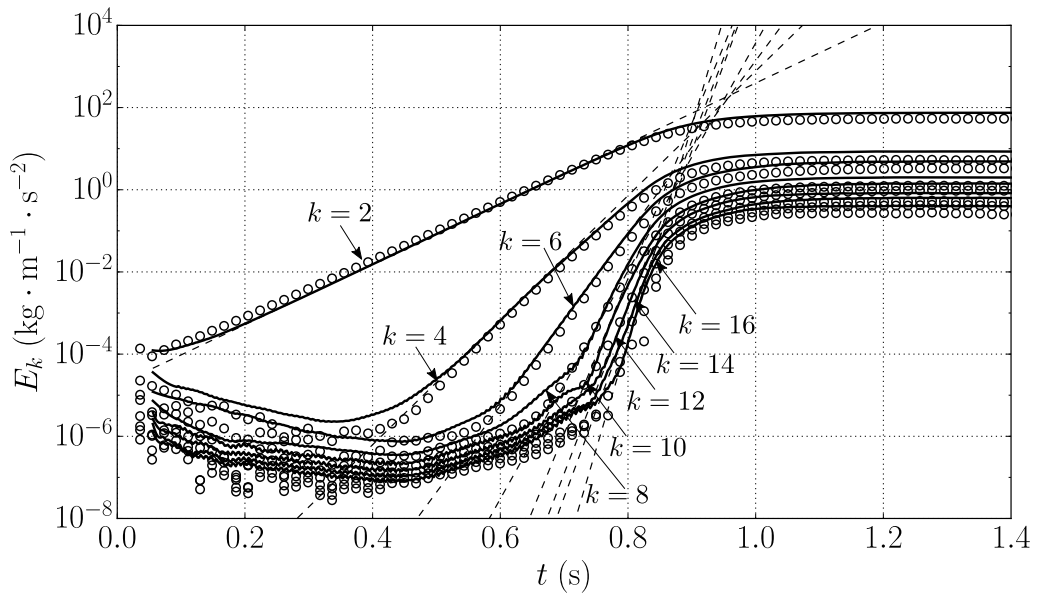}}
\caption{Comparison of evolution of the spectral energy density $E_k$ for the unstable mode and its first seven overtones as obtained from the signal shown in figure~\ref{fig: model_signal}. (--), Nonlinear model; ($\circ$), Navier-Stokes simulations; (- -), Linear interpolation of spectral energy density evolution in hierarchical spectral broadening regime.}
\label{fig: model_spectra}
\end{figure}

Figure~\ref{fig: model_spectra} shows the time evolution of the spectral energy density $E_k$  of the unstable mode (cf.~\eqref{eq:energy_spectral}) and its first seven overtones, as obtained from the nonlinear model discussed in \S~\ref{sec: NonlinearRegime_modelling}, and compares it to the results obtained from the fully compressible Navier-Stokes simulations for the signal shown in figure~\ref{fig: model_signal}. In the spectral broadening regime of thermoacoustic wave amplification, the growth of spectral energy density of the $k^{\mathrm{th}}$ harmonic obtained from linear interpolation is approximately $k\alpha_2/2$ (table~\ref{tab: spectral_growth}), where $\alpha_2=2\alpha$ (since $E\propto p'^2$) is the growth rate of the  spectral energy density of the unstable mode, i.e., $E_k\sim e^{k\alpha_2t/2}$. For instance, figure~\ref{fig: model_spectra} corresponds to the case $T_H = 450$\,K, for which the growth rate is $\alpha = 8.64\,\mathrm{s}^{-1}$; therefore, the growth rate of the spectral energy density is $\alpha_2 = 2\alpha=17.28\,\mathrm{s}^{-1}$ for the unstable mode.

Nonlinear energy cascade in the spectral space can be further explained using reduced order modelling. Assuming propagation of purely travelling waves in the system and eliminating $U'$ from equations~\eqref{eq: inviscid_p} and~\eqref{eq: inviscid_U}, the following  Burgers equation for the pressure fluctuations $p'$ is obtained:
\begin{equation}
\frac{\partial p'}{\partial t} - \frac{(\gamma+1)}{4\rho_0a_0}\frac{\partial p'^2}{\partial \xi} = \frac{\delta}{2}\frac{\partial^2 p'}{\partial \xi^2}, \label{eq: Burgers}
\end{equation}
where $\xi = a_0t - x$ is the travelling wave coordinate and $\delta$ is the axial dissipation coefficient given by
\begin{equation}
 \delta = \nu_0\left(\frac{4}{3} + \xi_B\right) + \frac{k}{\rho_0}\left(\frac{1}{C_v} - \frac{1}{C_p}\right). \label{eq: dissipation}
\end{equation}
In general, the wall-shear and the wall-heat flux expressions derived in~\eqref{eq: analytical_tau} and~\eqref{eq: analytical_q} can be used as forcing functions in the above Burgers equation. However, the abrupt area changes present in the setup under study make the generalization and further time domain simplification seemingly challenging. Hence, we seek to only qualitatively explain the temporal evolution of the nonlinear cascade utilizing the Burgers equation. Introducing thermoacoustic amplification by adding a simple linear forcing term in~\eqref{eq: Burgers} yields: 
\begin{equation}
 \frac{\partial p'}{\partial t} - \frac{(\gamma + 1)}{4\rho_0a_0}\frac{\partial p'^2}{\partial \xi} = \frac{\delta}{2}\frac{\partial^2 p' }{\partial \xi^2} + \alpha_{\mathrm{th}}p', \label{eq: Burgers_thermoacoustic}
\end{equation}
where $\alpha_{\mathrm{th}}$ accounts for the thermoacoustic growth rate. Substituting a Fourier expansion for acoustic pressure $p'$, namely,
\begin{equation}
 p' = \sum_{k}p_k(t)\sin\left(\frac{k\omega_0}{2a_0}\xi\right),~~\mathrm{where}~~k=2,4,6,8,\cdots
\end{equation}
the following modal evolution equation is obtained:
\begin{equation}
 \frac{dp_k}{dt} = \alpha_{\mathrm{th}}p_k + Q(p_{k}) - \frac{\delta}{16} \left(\frac{\omega_0}{a_0}\right)^2k^2p_k, \label{eq: CascadeFunction1}
\end{equation}
where 
\begin{equation}
 Q(p_k) = \frac{(\gamma+1)\omega_0}{8\rho_0a^2_0}\Bigg(\sum^{n\leq k-2}(k-n)p_np_{k-n} - k\sum_{n\geq k+2}p_np_{n-k}\Bigg), \label{eq: CascadeFunction}
\end{equation}
$\alpha_{\mathrm{th}}$ determines the rate of thermoacoustic amplification of the $k^{\mathrm{th}}$ mode ($\alpha_{\mathrm{th}} = 8.64\,\mathrm{s}^{-1}$ for $k=2$), and $\omega_0$ is the angular frequency of the unstable mode. 
$Q(p_k)$ is the nonlinear cascade function quantifying the scale-by-scale flux of energy in the spectral space from the unstable mode to the harmonics which are dissipated by the molecular diffusion effects (momentum and thermal diffusivity). The third term on the right-hand side of~\eqref{eq: CascadeFunction1} signifies dissipation of the $k^{\mathrm{th}}$ mode through molecular diffusion. Assuming that the subsequent overtones of the unstable mode are characterized by pressure amplitudes which are an order of magnitude smaller ($p_{k+2}/p_{k} \ll 1$), the modal growth rate~\eqref{eq: CascadeFunction1} can be approximated such that 
 \begin{equation}
  \frac{dp_2}{dt}\approx\alpha p_2,\quad\frac{dp_4}{dt}\approx\frac{(\gamma+1)}{4\rho_0a^2_0}\omega_0p^2_2,\quad\frac{dp_6}{dt}\approx\frac{3(\gamma + 1)}{4\rho_0a^2_0}\omega_0p_2p_4, \label{eq: overtones}
 \end{equation}
  and so on for higher harmonics. Equation~\eqref{eq: overtones} shows that the growth rates of the overtones of the unstable mode due to the energy cascade are proportional to the ratio of the oscillation frequencies, that is to say, $p_k \sim e^{k\alpha t/2}$ in the hierarchical spectral broadening regime. It is important to note that the growth of higher overtones is not strictly exponential since they are not subject to thermoacoustic instability. Energy is cascaded into the overtones of the thermoacoustically unstable mode via nonlinear energy cascade due to the high acoustic wave amplitude. Nonetheless, it is possible to assume modal growth for a small time interval in the spectral broadening regime to quantify the exponential growth rate for each harmonic (table~\ref{tab: spectral_growth}).

\begin{table}
\begin{tabular}{c}
\multicolumn{1}{c}{} \\
N-S\\
\\
\hline
Model\\
\\
\end{tabular}
\begin{tabular}{c|cccccccc}
$k$ & $2$ & $4$ & $6$ & $8$ & $10$ & $12$ & $14$  & $16$ \\
\hline
$\frac{\omega_k}{2\pi}~\mathrm{(Hz)}$  & 266.03 & 532.06 & 798.09 & 1064.12 & 1330.15 & 1596.18 & 1862.21 & 2128.24 \\
$\alpha_k~(\mathrm{s}^{-1})$ & 16.26 & 35.60 & 49.50 & 60.49 & 69.08 & 84.85 & 81.1 & 117.29 \\
\hline
$\frac{\omega_k}{2\pi}~\mathrm{(Hz)}$ & 265.25 & 530.50 & 795.76 & 1061.01 & 1326.26 & 1591.51 & 1856.76 & 2122.02 \\
$\alpha_k~(\mathrm{s}^{-1})$  & 16.92 & 34.89 & 48.36 & 63.85 & 86.96 & 93.21 & 97.04 & 120.87 \\
\hline
\end{tabular}\quad
 \caption{Frequency and growth rates of unstable mode and its first seven overtones as obtained from Navier-Stokes simulations (N-S) and time integration of the nonlinear model (Model).}
 \label{tab: spectral_growth}
\end{table}

\section{Thermoacoustically sustained shock waves}
\label{sec: LimitCycle}
The continued nonlinear energy cascade results in thermoacoustically sustained propagating shock waves at the limit cycle. While shock wave propagation outside the regenerator is quasi-planar, inside the regenerator, shock wave propagation is significantly affected by the wall-shear stresses and the wall-heat flux. In this section, time domain nonlinear model results are used to quantify the production of the energy density at the limit cycle (\S~\ref{sec: PhaseEvo}). Furthermore, utilizing the Navier-Stokes simulations and the nonlinear model results, a spectral scaling for the spectral energy density is derived; such a scaling, in turn, suggests the existence of \emph{thermoacoustic energy cascade} analogous to turbulent energy cascade (\S~\ref{sec: Spectral energy distribution}).

\subsection{Phase evolution}
\label{sec: PhaseEvo}
\begin{figure}
\centering
\includegraphics[width=0.85\textwidth]{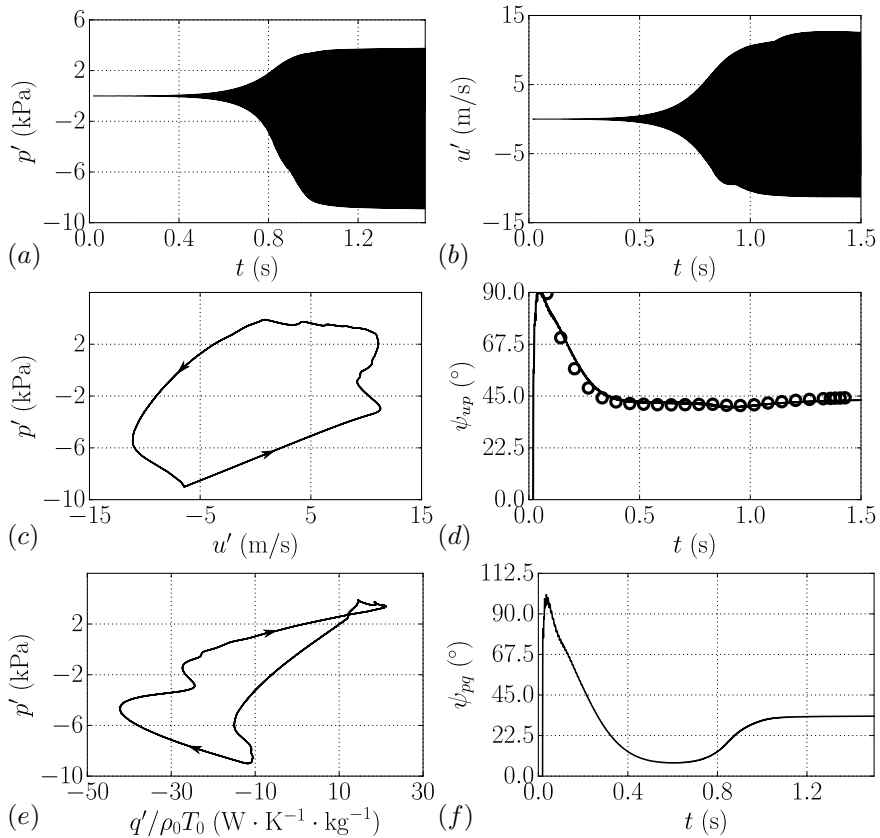}
\put(-327,215){$(a)$}
\put(-165,215){$(b)$}
\put(-327,110){$(c)$}
\put(-165,110){$(d)$}
\put(-327,5){$(e)$}
\put(-165,5){$(f)$}
\caption{Evolution of pressure perturbation $p'$  $(a)$, axial velocity perturbation $u'$ at $y=0$\,mm $(b)$, phase portrait in the $p'-u'$ plane at the limit cycle $(c)$, acoustic phasing $\psi_{pu}$ at $x=20$~mm inside the regenerator, phase portrait in the $p'-q'$ plane at the limit cycle $(e)$, and phase $\psi_{pq}$ evolution inside the regenerator at $x=20$\,mm $(f)$. $(d)$: (--), Nonlinear thermoacoustic model; ($\circ$), Navier-Stokes simulations.}
\label{fig: phasePU}
\end{figure}

As discussed in \S~\ref{sec: LinearRegime}, the cycle averaged production of the acoustic energy density $\overline{\mathcal{P}}$ is governed by the relative phasing between pressure and velocity  perturbations $p'$ and $u'$, respectively. Figures~\ref{fig: phasePU}$a - d$ show the phasing of $p'$ and $u'$ extracted at $x=20$\,mm, $y=0$\,mm and compare the evolution of the relative phase angle $\psi_{up}$ obtained from the time domain nonlinear model and the fully compressible Navier-Stokes simulations using the correlation function expression 
\begin{equation}
 \cos(\psi_{up}) = \frac{\int^t_0p'u'd\tau}{\sqrt{\int^t_0p'^2d\tau}\sqrt{\int^t_0u'^2d\tau}}. 
\end{equation}
In general, the above relation can be used to study the evolution of the phase difference between close to harmonically oscillating quantities~\citep{BalasubramanianS_JFM_2008}. The relative phase $\psi_{up}$ saturates at a value of approximately $40^{\circ}$, which is also the value in the harmonic growth regime indicating production of the energy density at the limit cycle inside the regenerator. However, due to large amplitude perturbations at the limit cycle, the production is attenuated due to the macrosonic thermoacoustic interaction term~\eqref{eq: MacrosonicThermoacousticInteraction} discussed in the previous section. The net production of the energy density inside the regenerator in the nonlinear regime can be approximated by combining equations~\eqref{eq: nonlinear_U} and~\eqref{eq: nonlinear_P_reduced} such that 
\begin{equation}
 \mathcal{P} - \mathcal{D} \approx\frac{p'}{h_b}\left(\frac{q'}{C_p\rho_0T_0}-\mathbb{Q}\right) + \frac{\tau'_wU'}{h^2_b},
 \label{eq: budget_nonlinear}
\end{equation}
which shows  a decrease, compared to the linear regime, due to the macrosonic thermoacoustic interaction term $\mathbb{Q}$. 

Furthermore, figure~\ref{fig: phasePU}$f$ shows that the phase angle $\psi_{pq}$ saturates at a value $\psi_{pq}\approx33.57^{\circ}$ compared to $\psi_{pq}\approx 7.41^{\circ}$ in the linear regime, thus showing the decrease in the production of the energy density at the limit cycle. However, since the phase angle $\psi_{pq}\neq90^{\circ}$ at the limit cycle, the production of acoustic energy density continues and shock waves are therefore sustained.

\subsection{Scales of acoustic spectral energy cascade}
\label{sec: Spectral energy distribution}
At the limit cycle, the energy of the unstable mode continues to increase and further cascades into higher harmonics on account of the nonlinear wave propagation. Higher harmonics have correspondingly shorter wavelengths due to which gradients in the longitudinal direction $x$ become large. Through the bulk viscosity and the thermal conductivity, the energy density is dissipated at higher harmonics, thus establishing a steady flow of energy from the unstable mode to the higher harmonics. The distribution of the spectral energy density in the harmonics can be derived utilizing an energy cascade modelling, analogous to the turbulent energy cascade~\citep{pope2000turbulent,Nazarenko_2011_WT}. Assuming travelling wave propagation, the total energy density of the $k^{\mathrm{th}}$ harmonic associated to planar wave propagation, $E^{(1\mathrm{D})}_k$, can be defined as 
\begin{equation}
 E^{(1\mathrm{D})} = \int^{\frac{2\pi a_0}{\omega_0}}_0E\,d\xi = \sum_k E^{(1\mathrm{D})}_k, \quad E^{(1\mathrm{D})}_k = \frac{\pi a_0}{\omega_0}E_k, \label{eq: TotalEnergy}
\end{equation}
where $E$ is the squared $L_2$ norm defined in~\eqref{eq:energy_time} where $p'$ and $u'$ are defined as a function of the travelling wave coordinate $\xi$. At the limit cycle, the rate of cascade of energy $\mathcal{E}$ in the spectral space balances the thermoacoustic wave amplification. As a result, higher harmonics (namely, the overtones of the unstable mode) are generated and later  dissipated by molecular dissipation, as indicated by the presence of  the molecular dissipation factor $\delta$ in~\eqref{eq: dissipation}. Thus, $\mathcal{E}$ is purely governed by thermoacoustic wave amplification at smaller harmonics and viscous dissipation at higher harmonics and scales as 
\begin{equation} \label{eq:dissipation_rate}
 \frac{\mathcal{E}}{\alpha^2_{\mathrm{eff}}\delta} = \mathrm{const.},
\end{equation}
where $\alpha_{\mathrm{eff}}$ is the effective energy amplification rate at the limit cycle.
\begin{figure}
\centering
  \centerline{\includegraphics[width=1.0\textwidth]{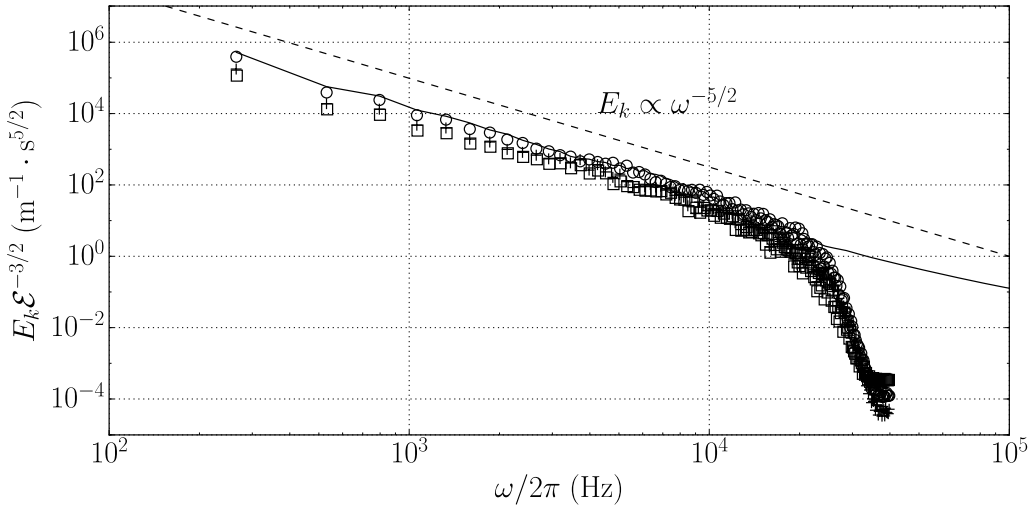}}
  \caption{Scaled spectral energy density $E_k\mathcal{E}^{-3/2}$ at the limit cycle against frequency of the harmonics. (--), Time domain nonlinear model at $T_H=450~$; Navier-Stokes simulations at ($\circ$), $T_H=450~$K; ($+$), $T_H = 500$\,K; ($\square$),  $T_H = 550$\,K; (- -) compares the variation of the energy with harmonic frequency with the power law derived in~\eqref{eq: SpectralScaling}.}
\label{fig: pressure_spectralDist}
\end{figure}
Based on the macrosonic thermoacoustic interaction, $\alpha_{\mathrm{eff}}$ can be estimated as
\begin{equation}
\alpha_{\mathrm{eff}}\approx(1-\gamma |p'|/p_0)\alpha,
\end{equation}
 where $\alpha$ is the growth rate of the unstable mode in the linear regime, and $|p'|$ is the amplitude of the pressure fluctuation at the limit cycle. Assuming that the total energy per unit mass $E^{(1\mathrm{D})}_k/\rho_0$ only depends on the rate of energy cascade $\mathcal{E}$ and on the angular frequency of the harmonic $\omega_k$, the following scaling is obtained: 
\begin{equation}
 \frac{E^{(1\mathrm{D})}_k \omega^{5/2}_k}{\rho_0\mathcal{E}^{3/2}} = \mathrm{const.}. \label{eq: SpectralScaling0}
\end{equation}
Substituting into~\eqref{eq: TotalEnergy} and eliminating $\rho_0$, the scaling for the spectral energy density $E_k$ remains the same such that 
\begin{equation}
 \frac{E_k\omega^{5/2}_k}{\mathcal{E}^{3/2}} = \mathrm{const.}. \label{eq: SpectralScaling}
\end{equation}
The scaling derived in~\eqref{eq: SpectralScaling} shows that the distribution of the energy density at limit cycle in the spectral space decays as $\omega^{-5/2}_k$---where $\omega_k$ is the frequency of the harmonic---as shown in figure~\ref{fig: pressure_spectralDist}. Moreover, in the spectral broadening regime, the energy density in the spectral space varies with the effective growth rate as $\alpha^{3}_{\mathrm{eff}}$ (figure~\ref{fig:intro_spectral}) which, in turn, allows the comparison among cases with varying hot side temperature $T_H$. Such a scaling arises purely from the mechanism of the nonlinear saturation resulting from the spectral energy cascade. It is however challenging to accurately estimate the acoustic energy production (quantified by $\alpha^{3}_{\mathrm{eff}}$) at the limit cycle utilizing the time domain nonlinear model, or even the fully compressible Navier-Stokes simulations, due to the shock capturing artificial viscosity that has to be added to ensure numerical stability in the presence of shocks. Overall, the spectral cascade of the energy density is balanced by the thermoacoustic wave amplification and the viscous dissipation, i.e., the shock waves are thermoacoustically sustained. However,  as shown previously~\citep{GuptaLS_SCITECH_2017}, at the location of the steepest gradient in pressure, the wall-heat flux from the acoustic field inside the isothermal walls and wall shear stress are maximum inside the regenerator. Thus, the propagation of sharp gradients in the acoustic field $p'$ and $U'$ inside the regenerator are \emph{counteracted} by the wall-shear stresses and the wall-heat flux resulting in smoothing of shock waves. Consequently, the scaling argument~\eqref{eq: SpectralScaling} can be further improved accounting for propagation of very large harmonics inside the regenerator.


\section{Conclusions}

The linear and nonlinear regimes of thermoacoustic wave amplification have been modelled up to the formation of shock waves in a minimal unit looped resonator with the support of high-fidelity fully compressible Navier-Stokes simulations. The computational setup is inspired by the experimental investigations conducted by~\citet{Yazaki_PhysRevLet_1998} and geometrically optimized to maximize growth rates for the quasi-travelling wave mode. Three regimes of thermoacoustic wave amplification have been identified: ($i$) a monochromatic or modal growth regime, ($ii$) a hierarchical spectral broadening or nonlinear growth regime and ($iii$) a shock-dominated limit cycle. The modal growth regime is characterized by exponential amplification of thermoacoustically unstable modes. 
An acoustic energy budget formulation yielding a closed form expression of the Rayleigh index has been developed and the effect of variations in geometry and hot-to-cold temperature ratios on the thermoacoustic growth rates have been elucidated. The limit cycle regime exhibits many features of Kolmogorov's equilibrium turbulence, where energy, steadily injected at the integral length scale (wavelength of the second-mode harmonic), cascades towards higher wave numbers via inviscid mechanisms (wave steepening) and is finally dissipated at the Kolmogorov's length scale (of the order of the shock thickness). A grid sensitivity analysis has been carried out at the limit cycle to ensure that the entropy jump across the captured shock waves is grid convergent, hence assuring the same fidelity typically attributed to direct numerical simulations of turbulent flows, with the caveat that shocks in the present study are not fully resolved by the computational mesh.

The existence of an equilibrium \emph{thermoacoustic energy cascade} has thus been shown. The spectral energy density at the limit cycle, in particular, has been found to decay as $\omega^{-5/2}$ in  spectral space, the relevant intensity scaling with growth rate $\alpha$ as $\alpha^3$. Such findings are confirmed by a novel theoretical framework to model  thermoacoustic nonlinearities, which has lead to the formulation of a one-dimensional time-domain nonlinear acoustic model. The model is correct up to second order in the perturbation variables and addresses the fundamental problem of high amplitude wave propagation in the presence of wall-shear and wall-heat flux, accounting for thermodynamic nonlinearities such as the second-order interactions between the pressure fluctuations and the wall-heat flux, namely the macrosonic thermoacoustic interaction term. The model also confirms the dynamics of energy transfer across scales in the nonlinear growth regime: the growth of higher harmonics is hierarchical in nature and higher harmonics are amplified at faster rates. in particular the $k^{\mathrm{th}}$ harmonic grows at the rate of $\alpha k/2$, where $\alpha$ is the rate of growth of energy in the unstable mode. 

\section*{Acknowledgments}
Prateek Gupta acknowledges the support of the Lynn Fellowship at Purdue University, Ford Motor Company and Rolls-Royce Corporation (Indianapolis). 
The use of the spectral difference solver originally developed by Antony Jameson's group at Stanford University, and financial support of Guido Lodato by CNRS under the INSA Turbulence and Simulation Chair are also gratefully acknowledged.
Computations have been run on the high-performance computing resources provided by the Rosen Center for Advanced Computing (RCAC) at Purdue University and by the Institut du D\'eveloppement et des Ressources en Informatique Scientifique (IDRIS-CNRS) under the allocation i2015-2a7361 and of the Haute Normandie Computing center CRIANN. 

\appendix
\section{Experimental Validation of linear model}
\label{sec: app0}
\begin{figure}
\includegraphics[width=0.95\textwidth]{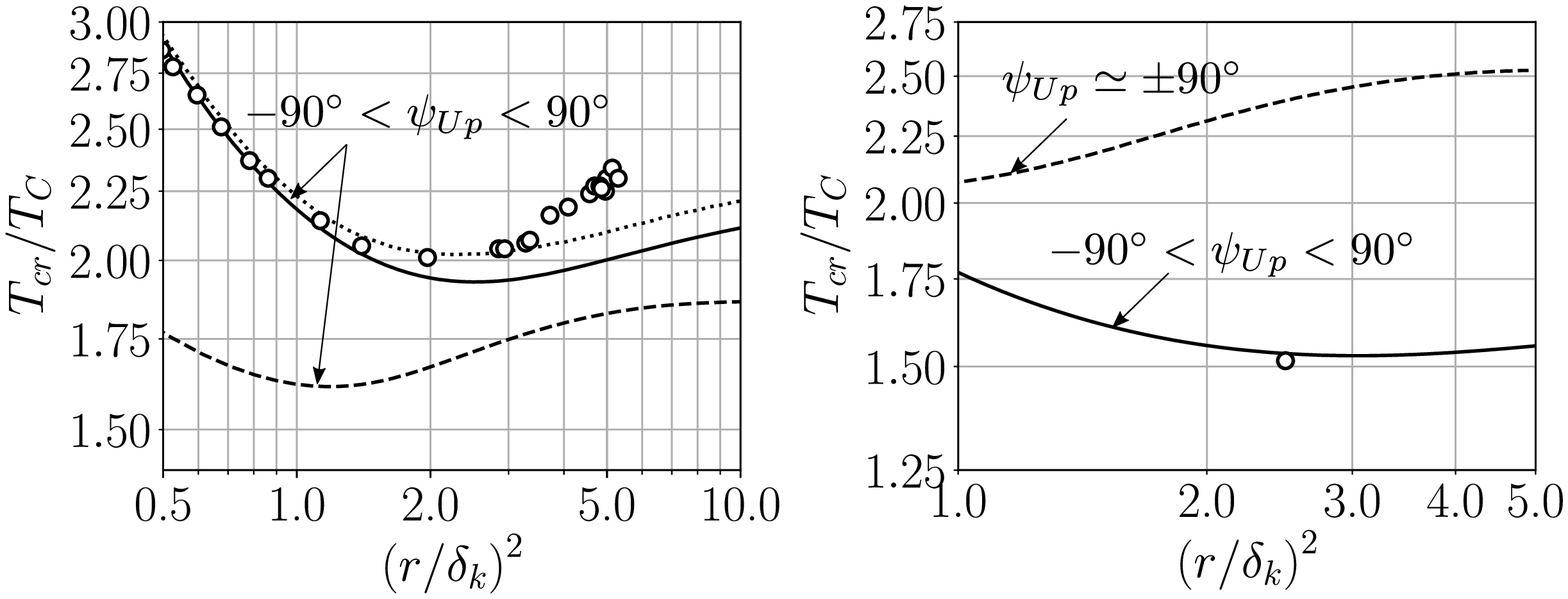}
\put(-355,7){$(a)$}
\put(-180,7){$(b)$}
\caption{Neutral stability curves obtained from linear analysis (\S~\ref{sec: Linear}), ~\cite{Yazaki_PhysRevLet_1998}'s data ($a$), and~\cite{BiwaTY_JAcousSocAm_2011}'s reported instability limit ($b$). $r$ is the hydraulic radius of regenerator pore (square cross-section). ($a$): (--), Y-1998 in 3D ; (- -), 3D minimal unit corresponding to Y-1998; ($\cdot \cdot\cdot$),~\cite{guedra_use_2012}'s results; ($\circ$),~\cite{Yazaki_PhysRevLet_1998}'s reported data. ($b$): (--), B-2011 in 3D (with $\ell_{buffer}=0.4~$m); (- -), 3D minimal unit corresponding to B-2011 without buffer length ; ($\circ$), reported instability limit.}
\label{fig: ExpValid}
\end{figure}

Figure~\ref{fig: ExpValid}$a$ shows the comparison between the neutral stability curve evaluated from the linear eigenvalue analysis discussed in \S~\ref{sec: Linear}, experimental data obtained by~\cite{Yazaki_PhysRevLet_1998} (setup referenced as Y-1998), and the numerical predictions obtained by~\cite{guedra_use_2012}. The lengths of smaller cross-section ducts ($\ell_a$ and $\ell_c$ in figure~\ref{fig:computational_setup}) have been estimated from the laser Doppler velocimetry (LDV) data reported by~\cite{Yazaki_PhysRevLet_1998} ($\ell_a=0.153$ m and $\ell_c=0.5$ m). The rest of the missing geometrical details have been taken from~\cite{guedra_use_2012}. 
The looped geometry of ~\cite{BiwaTY_JAcousSocAm_2011} (setup referenced as B-2011), is shown to be thermoacoustically unstable at the reported critical $T_{H}$ value, ($T_{cr}/T_{C}=1.52$) only if a mean negative temperature gradient from $\ell_b<x<\ell_{\mathrm{buffer}}+\ell_b$, is assumed in the linear analysis. Moreover, the unstable mode is a quasi-travelling wave ($-90^{\circ}<\psi_{Up}<90^{\circ}$). However, by considering an abrupt temperature change from $T_{H}$ to ambient temperature $T_{C}$, the ratio $T_{cr}/T_{C}$ is very large ($\sim$2.3) and the unstable mode is a standing wave ($\psi_{Up}\simeq \pm 90^{\circ}$). Since the nonlinear cascade leading to shock formation is inhibited by the standing wave phasing compared to the quasi-travelling wave phasing~\citep{BiwaEtAl_JASA_2014}, a geometry similar to Y-1998 has been chosen in the current study which exhibits quasi-travelling wave unstable mode due to a wider duct section.

\section{Scaling of governing equations}
\label{sec: app1}
The governing equations for compressible flows in two dimensions read: 
\begin{subequations}
 \begin{gather}
  \frac{\partial \rho}{\partial t} + \frac{\partial (\rho u)}{\partial x} + \frac{\partial (\rho v)}{\partial y} = 0,\\
  \rho\left(\frac{\partial u}{\partial t} + u\frac{\partial u}{\partial x} + v\frac{\partial u}{\partial y}\right) = -\frac{\partial p}{\partial x} + \frac{\partial}{\partial x}\left[\mu\left(\xi_B + \frac{4}{3}\right)\frac{\partial u}{\partial x}\right] + \frac{\partial }{\partial y}\left(\mu\frac{\partial u}{\partial y}\right),\\
  \rho\left(\frac{\partial v}{\partial t} + u\frac{\partial v}{\partial x} + v\frac{\partial v}{\partial y}\right) = -\frac{\partial p}{\partial y} + \frac{\partial}{\partial y}\left[\mu\left(\xi_B + \frac{4}{3}\right)\frac{\partial v}{\partial y}\right] + \frac{\partial}{\partial x}\left(\mu\frac{\partial v}{\partial x}\right),\\
  \rho T \left(\frac{\partial s}{\partial t} + u\frac{\partial s}{\partial x} + v\frac{\partial s}{\partial y}\right) = \frac{\partial }{\partial y}\left(k\frac{\partial T}{\partial y}\right) + \frac{\partial }{\partial x}\left(k\frac{\partial T}{\partial x}\right) + \Phi,
 \end{gather}
\end{subequations}
where $\Phi$ accounts for entropy generation due to viscous gradients. These equations are written in terms of perturbations, denoted by primed symbols, in the relevant variables $u$, $v$, $\rho$, $p$, $T$, and $~s$.
It shall be noted that, for a nonlinear acoustic field of velocity amplitude scale $\mathcal{U}$, the acoustic Mach number $\mathcal{M}=\mathcal{U}/a_0$ is $\mathcal{O}(10^{-1})$ and the aspect ratio of the regenerator $h_b/l_b$ is $\mathcal{O}(10^{-2})$. As a result, the perturbations in the $y$ component of velocity (namely, $v'/u' \sim h_b/l_b$) from the momentum and entropy equations can be neglected and the following equations, including second order terms  in $\mathcal{M}$, are obtained:
\begin{subequations}
\begin{gather}
\frac{\partial \rho'}{\partial t} + \rho_0 \frac{\partial u'}{\partial x} + u'\frac{d\rho_0}{dx}  + \rho_0\frac{\partial v'}{\partial y} = \left[-\rho'\frac{\partial u'}{\partial x} - u'\frac{\partial \rho'}{\partial x}\right],\label{eq: app_mass}\\
\frac{\partial u'}{\partial t} + \frac{1}{\rho_0}\frac{\partial p'}{\partial x} - \frac{1}{\rho_0}\frac{\partial}{\partial y}\left(\mu\frac{\partial u'}{\partial y}\right) - \frac{1}{\rho_0}\frac{\partial }{\partial x}\left[\mu\left(\xi_B + \frac{4}{3}\right)\frac{\partial u'}{\partial x}\right] =   \left[-\frac{\rho'}{\rho_0}\frac{\partial u'}{\partial t} - \frac{1}{2}\frac{\partial u'^2}{\partial x}\right],\label{eq: app_mom}\\
\frac{\partial s'}{\partial t} + u'\frac{ds_0}{dx} - \frac{R}{p_0}\frac{\partial}{\partial y}\left(k\frac{\partial T'}{\partial y}\right) - \frac{R}{p_0}\frac{\partial}{\partial x}\left(k\frac{\partial T'}{\partial x}\right) = \left[-\frac{p'}{p_0}\left(\frac{\partial s'}{\partial t} + u'\frac{ds_0}{dx}\right) - u'\frac{\partial s'}{\partial x}+\Phi\right].\label{eq: app_ent}
\end{gather}
\end{subequations}
In the above equations, the subscript $0$ denotes the base state. Also, the terms on the left-hand side of the equations are linear in the perturbation variables, whereas the terms on the right-hand side are non-linear. Assuming $\omega^{-1}$ as the characteristic time scale of the acoustic field,  the viscous dissipation terms on the left-hand side of~\eqref{eq: app_mom} and~\eqref{eq: app_ent}, and the entropy generation due to viscous gradients $\Phi$ scale relatively, as
\begin{eqnarray}
 \frac{\left\lvert \frac{1}{\rho_0}\frac{\partial}{\partial y}\left(\mu\frac{\partial u'}{\partial y}\right) \right\lvert}{\left\lvert \frac{\partial u'}{\partial t} \right\lvert} \sim \left(\frac{\delta_\nu}{h_b}\right)^2 \left(1 + \mathcal{M}\right),\quad \frac{\left\lvert \mu \left(\frac{\partial u'}{\partial y}\right)^2\right\lvert}{\rho T \frac{\partial s}{\partial t}} \sim \left(\frac{\delta_\nu}{h_b}\right)^2 \mathcal{M},\\ \frac{\left\lvert \frac{1}{\rho_0}\frac{\partial }{\partial x}\left[\mu\left(\xi_B + \frac{4}{3}\right)\frac{\partial u'}{\partial x}\right] \right\lvert}{\left\lvert \frac{\partial u'}{\partial t} \right\lvert} \sim \left(\frac{\delta_\nu}{h_b}\right)^2 \left(\frac{h_b\omega}{a_0}\right)^2 \left(1 + \mathcal{M}\right),
\end{eqnarray}
where the following scaling relations have been used:
\begin{equation}
 \mu \sim \mu_0(1 + \mathcal{M}),\quad \delta_{\nu} \sim \sqrt{\frac{\nu_0}{\omega}}. 
\end{equation}
Due to the nonlinear energy cascade, higher harmonics are generated, as a result of which, the characteristic time scale of the acoustic field $\omega^{-1}$ decreases. Consequently, the characteristic Stokes layer thickness, $\delta_\nu$, decreases and the ratio $h_b\omega/a_0$ increases. Accordingly, the terms which scale as $\mathcal{O}((\delta_\nu/h_b)^2\mathcal{M})$ can be dropped and terms corresponding to the local perturbations in viscosity and thermal conductivity, as well as the entropy generation term due to viscous gradients, are neglected. Finally, the equations governing the spatio-temporal evolution of the acoustic field, correct up to the second order, are given as 
\begin{subequations}
\begin{gather}
\frac{\partial \rho'}{\partial t} + \rho_0 \frac{\partial u'}{\partial x} + u'\frac{d\rho_0}{dx} + \rho_0\frac{\partial v'}{\partial y} = \left[-\rho'\frac{\partial u'}{\partial x} - u'\frac{\partial \rho'}{\partial x}\right],\label{eq: 2_massApp}\\
\frac{\partial u'}{\partial t} + \frac{1}{\rho_0}\frac{\partial p'}{\partial x} - \nu_0 \frac{\partial^2 u'}{\partial y^2} - \frac{1}{\rho_0}\frac{\partial }{\partial x}\left[\mu_0\left(\xi_B + \frac{4}{3}\right)\frac{\partial u'}{\partial x}\right] =  \left[-\frac{\rho'}{\rho_0}\frac{\partial u'}{\partial t} - \frac{1}{2}\frac{\partial u'^2}{\partial x}\right],\label{eq: 2_momApp}\\
\frac{\partial s'}{\partial t} + u'\frac{ds_0}{dx} - \frac{Rk_0}{p_0}\frac{\partial^2 T'}{\partial y^2} - \frac{R}{p_0}\frac{\partial}{\partial x}\left(k_0\frac{\partial T'}{\partial x}\right) = \left[-\frac{p'}{p_0}\left(\frac{\partial s'}{\partial t} + u'\frac{ds_0}{dx}\right) - u'\frac{\partial s'}{\partial x}\right], \label{eq: 2_entApp}
\end{gather}
\end{subequations}

\section{Time domain model and convergence}
\label{sec: app2}
Integrating~\eqref{eq: 2_massApp} and~\eqref{eq: 2_entApp} in $y$, the following equations are obtained:
\begin{equation}
\frac{\partial \tilde{\rho}^{(2)}}{\partial t} + \rho_0\frac{\partial U'^{(1)}}{\partial x} + U'^{(1)}\frac{d\rho_0}{dx} = \int^{h_b/2}_{-h_b/2}\left[-\rho'^{(1)}\frac{\partial u'^{(1)}}{\partial x} - u'^{(1)}\frac{\partial \rho'^{(1)}}{\partial x}\right]dy,\label{eq: ContApp}
\end{equation}
\begin{equation}
  \frac{\partial S'^{(2)}}{\partial t} + U'^{(1)}\frac{ds_0}{dx} = \left[1-\frac{p'^{(1)}}{p_0}\right]\frac{q'^{(1)}}{\rho_0T_0} + q_2 + \frac{R}{p_0}\frac{\partial }{\partial x}\left[\frac{k_0T_0}{C_p}\frac{\partial S'^{(1)}}{\partial x}\right] -
\int^{h_b/2}_{-h_b/2}u'^{(1)}\frac{\partial s'^{(1)}}{\partial x}dy\label{eq: EntApp},
\end{equation}
where,
\begin{gather}
 \tilde{\rho}^{(2)} = \int^{h_b/2}_{-h_b/2}\rho^{(2)} dy,\quad S'^{(2)}= \int^{h_b/2}_{-h_b/2}s'^{(2)} dy, \\
 q'^{(1)} = \frac{2\nu_0\rho_0 T_0}{Pr}\left.\frac{\partial s'^{(1)}}{\partial y}\right|_{y=h_b/2}.
\end{gather}
Also, the density constitutive equation, up to second order, is given by 
\begin{equation}
 \rho'^{(2)} = \alpha_s p'^{(2)} + \alpha_ps'^{(2)} + \frac{1}{2}\left(\beta_s p'^{(1)2} + \beta_p s'^{(1)2} + 2\beta_{sp} s'^{(1)}p'^{(1)}\right), \label{eq: ConstitutiveApp}
\end{equation}
and the entropy perturbations $s'^{(1)}$, up to first order, are 
\begin{equation}
 s'^{(1)} = -\frac{R}{p_0}p'^{(1)} + \sum_{j=0}^{\infty}\check{s}_j(x,t)\cos \left(\zeta_j y\right).\label{eq: EntropyModesApp}
\end{equation}
Notice that the superscripts $(1)$ and $(2)$ are dropped hereafter for convenience of notation. Integrating equation~\eqref{eq: ConstitutiveApp} along $y$, differentiating in time, and combining with~\eqref{eq: ContApp} and~\eqref{eq: EntApp} to eliminate $\tilde{\rho}$, the following equation is obtained: 
\begin{equation}
\frac{\partial p'}{\partial t} + \frac{\rho_0a^2_0}{h_b}\frac{\partial U'}{\partial x}=\frac{\rho_0a^2_0}{h_b}\left(\frac{q'}{C_p\rho_0T_0}+q_2  + \mathbb{T} -\mathbb{Q} +  \mathbb{D}_s\right) - \mathbb{C}, \label{eq: nonlinear_P_App}
\end{equation}
where 
\begin{equation}
 q' = \frac{2\nu_0\rho_0 T_0}{Pr}\sum^{\infty}_{j=0}(-1)^{j+1}\check{s}_j(x,t)\zeta_j,\quad\mathbb{Q} = \frac{\gamma p'q'}{C_p p_0\rho_0T_0},
 \label{eq: heat_flux}
\end{equation}
and,
\begin{subequations}
\begin{gather}
q_2 = \frac{h_b\nu_0}{C_p Pr}\sum^{\infty}_{j=0}\left(\zeta_j \check{s}_j\right)^2,\\
\mathbb{T} = \frac{\left(\gamma-1\right)}{p_0}p'\frac{\partial U'}{\partial x} + \frac{1}{C_pa^2_0}\frac{\partial}{\partial t}\Bigg[\frac{p'}{p_0}\sum_{j=0}^{\infty}(-1)^{j+1}\frac{2\check{s}_j}{\zeta_j}\Bigg]-\frac{h_b}{4C^2_pa^2_0}\frac{\partial}{\partial t}\sum^{\infty}_{j=0}s^2_j,\\
\mathbb{C} = -\frac{\gamma - 1}{a^2_0}U'\frac{\partial p'}{\partial x} -\gamma\frac{\partial}{\partial x}\left(\frac{p'U'}{a^2_0}\right)- \frac{\rho_0 h_b}{2C_p}\sum^{\infty}_{j=0}\check{u}_j\frac{\partial \check{s}_j}{\partial x} + \frac{h_b}{2}\frac{\partial}{\partial x}\Bigg(\frac{\rho_0}{C_p}\sum^{\infty}_{j=0}\check{u}_j\check{s}_j\Bigg),\\
\mathbb{D}_s = \frac{1}{Pr}\frac{\partial}{\partial x}\Bigg[\nu_0\Bigg(\sum^{\infty}_{j=0}(-1)^{j}\frac{2}{\zeta_j}\frac{\partial \check{s}_j}{\partial x} + \frac{h_b R}{p_0}\frac{\partial p'}{\partial x}\Bigg)\Bigg].
\end{gather}
\label{eq: nonlinearities_App}
\end{subequations}
The above equations account for thermodynamic, as well as, convective nonlinearities, with $\mathbb{D}_s$ denoting the axial conduction term. The treatment of quadratic nonlinearities gets significantly simplified  due to the Fourier expansions of the viscous and entropic modes. For simplicity, only the nonlinear macrosonic thermoacoustic interaction $\mathbb{Q}$ is retained  in the present one-dimensional computations. However, the complete model equations~\eqref{eq: nonlinear_P_App}--\eqref{eq: nonlinearities_App} should be considered in the case of relatively large regenerators. 

In order to show the convergence of the viscous and entropic modes, $\check{u}_j$ and  $\check{s}_j$, respectively, \eqref{eq: u_modes} and~\eqref{eq: s_modes} are pre-multiplied by $(-1)^{j+1}\zeta_j$ and summed over $j$ to obtain: 
\begin{gather}
 \frac{\partial \tau'_w}{\partial t} + 2\nu^2_0\sum^{N}_{j=0}(-1)^{j+1}\check{u}_j\zeta^3_j = N\frac{4\nu_0}{h\rho_0}\frac{\partial p'}{\partial x},\label{eq: shear_conv1}\\
 \frac{1}{\rho_0T_0}\frac{\partial q'}{\partial t} + \frac{\tau'_w}{Pr}\frac{ds_0}{dx} + 2\left(\frac{\nu_0}{Pr}\right)^2\sum^{N}_{j=0}(-1)^{j+1}\check{s}_j\zeta^3_j = -N\frac{4\nu_0}{h_bPr\rho_0 T_0}\left(\frac{\partial p'}{\partial t}\right). \label{eq: heat_conv1}
\end{gather}
For the wall-shear $\tau'_w$ and the wall-heat flux $q'$ to converge, \eqref{eq: shear_conv1} and~\eqref{eq: heat_conv1} must yield the same values of $q'$ and $\tau'_w$ for $N$ and $N+1$ in the limit of $N\rightarrow \infty$. Hence, the following conditions are obtained:  
\begin{equation}
 \lim_{N\to\infty} |u_N| = \frac{2}{\zeta^3_N}\frac{1}{h_b\rho_0\nu_0}\frac{\partial p'}{\partial x}, \quad \lim_{N\to\infty} |s_N| = \frac{2}{\zeta^3_N}\frac{Pr}{h_b\rho_0\nu_0 T_0}\frac{\partial p'}{\partial t}. \label{eq: modes_convergence}
\end{equation}
The above relations show that, in order for convergence to be ensured, the magnitudes of the viscous modes $\check{u}_j$ and $\check{s}_j$ must decay as $\zeta^{-3}_j$ for large values of the index $j$. The relaxation functional used in~\eqref{eq: analytical_un} and~\eqref{eq: analytical_sn} for very large $j$ yields: 
\begin{equation}
 \int^{t}_{-\infty}e^{-\frac{t-\eta}{\tau_j}}\phi(x,\eta)d\eta \approx \tau_j\phi(x,t), \label{eq: approx_kernel}
\end{equation}
for some function $\phi(x,t)$. After substituting the above approximation in~\eqref{eq: analytical_un} and~\eqref{eq: analytical_sn}, it is  shown that the magnitudes of the viscous modes $\check{u}_j$ and $\check{s}_j$ decay as $\zeta^{-3}_j$ in the limit of $j$ such that the approximation in~\eqref{eq: approx_kernel} holds valid, which, in turn,  guarantees the convergence of the infinite series approximation for the  viscous and entropic modes. 

\section{Windowed shock capturing}
\label{sec: app3}
At the limit cycle, the perturbation fields exhibit shock wave propagation causing the formation of large gradients at the limit of grid resolution. Spatially windowed Legendre polynomial expansions are therefore used in order to add an artificial viscosity term to ensure a proper resolution of longitudinal gradients in the model equations. The implemented  strategy, which is briefly summarized below, is similar to the one proposed by~\citet{persson:06} for discontinuous Galerkin methods.
For a spatial window of size $N_p$, the local pressure field is reconstructed using  Legendre polynomial basis expansions up to orders $N_p$ and $N_p-1$, such that 
\begin{equation}
 p'=\sum^{N_p}_{n=0}p_n\psi_n,\quad \tilde{p}' = \sum^{N_p-1}_{n=0}p_n\psi_n,
\end{equation}
where $\psi_n$ denotes the Legendre polynomial of order $n$ and $p_n$ is the relevant $n^{\mathrm{th}}$ mode. 
A modal smoothness indicator $S_e$ is hence constructed from the ratio of the inner products of  $p'-\tilde{p}'$ and $p'$, such as to detect the onset of excessively high high-frequency modes, which is typical of insufficiently unresolved signals:
\begin{equation}
 S_e = \frac{\langle p'-\tilde{p}',p'-\tilde{p}'\rangle}{\langle p',p' \rangle}.
\end{equation}
The sensor is then used to trigger the shock capturing artificial diffusivity $\epsilon_e$, which is evaluated as 
\[
\epsilon_e=
\begin{cases}
0 &\text{if } s_e < s_0  ,\\
\epsilon_0\sin\left(\frac{\pi(s_e-s_0)}{2\kappa}\right) &\text{if } s_0<s_e<s_0+\kappa,\\
\epsilon_0 &\text{if } s_e>s_0.
\end{cases}
\]
where $s_e = \log S_e$ and the corresponding parameters are  $s_0 = 3.2$, $\kappa = 0.9(\mathrm{max}(s_e) - s_0)$, and $\epsilon_0 = 100\Delta x/N_p$, $\Delta x$ being the grid spacing of the discretized one-dimensional domain.

Finally, the artificial viscosity augmented equations to capture the nonlinear acoustic fields become: 
\begin{gather}
   \frac{\partial p'}{\partial t} = -\frac{\gamma p_0}{h}\left(1 + \frac{1+\gamma}{\gamma}\frac{p'}{p_0}\right)\frac{\partial U'}{\partial x} + \left[\epsilon_{e}+\frac{k}{\rho_0}\left(\frac{1}{C_v} - \frac{1}{C_p}\right)\right]\frac{\partial^2 p'}{\partial x^2},\label{eq: inviscid_pApp}\\
   \frac{\partial U'}{\partial t} = -\frac{h}{\rho_0}\frac{\partial p'}{\partial x} + \left[\epsilon_e Pr+\nu_0\left(\xi_B + \frac{4}{3}\right)\right]\frac{\partial^2 U'}{\partial x^2}.\label{eq: inviscid_UApp}
\end{gather}

\bibliographystyle{jfm}
\bibliography{references}
\end{document}